\documentclass[a4paper,11pt]{article}
\pdfoutput=1 
\usepackage{jcappub, bm, color} 
\usepackage{amssymb,amsfonts,slashed,amsthm,amsmath,graphicx, soul, empheq}
\usepackage[caption=false]{subfig}
\newcommand{\eV}{ \ {\rm eV} }
\newcommand{\KeV}{ \ {\rm keV} }
\newcommand{\MeV}{\  {\rm MeV} }
\newcommand{\GeV}{\,{\rm GeV}}

\newcommand{\bea}{\begin{array}}
\newcommand{\eea}{\end{array}}
\newcommand{\beq}{\begin{eqnarray}}
\newcommand{\eeq}{\end{eqnarray}}

\newcommand{\Mpl}{M_{\rm Pl}}

\def\lrfp#1#2#3{ \left(\frac{#1}{#2} \right)^{#3}}

\title{
Stochastic Axion Dark Matter in Axion Landscape
}

\author{
Shota Nakagawa$^{\spadesuit }$
}
\affiliation{$^\spadesuit$ Department of Physics, Tohoku University, 
Sendai, Miyagi 980-8578, Japan} 

\author{
Fuminobu Takahashi$^{\spadesuit \diamondsuit}$
}

\affiliation{$^\diamondsuit$ Kavli IPMU (WPI), UTIAS, 
The University of Tokyo, 
Kashiwa, Chiba 277-8583, Japan}

\author{
Wen Yin$^{\heartsuit}$
}
\affiliation{
$^{\heartsuit}$ Department of Physics, Korea Advanced Institute of Science and Technology, Daejeon 34141, Korea
}

\abstract{
We study the stochastic axion dark matter scenario in the axion landscape, where one of the axions 
is light and stable and therefore explains dark matter.
If the axion mass at the potential is a typical value of the curvature along the direction, the potential
can be well approximated by a quadratic mass term. On the other hand, 
if the axion mass happens to be suppressed in the vicinity of the minimum, the potential may be approximated 
by a quartic potential plus a suppressed quadratic one, for which the initial angle, and thus the axion abundance, 
can be significantly suppressed  compared to the quadratic case. We delineate the viable parameter region
by taking account of various observational constraints, and find that a broader range of the inflation scale is allowed. 
Also, if the curvature of the potential is suppressed over a certain range of the 
potential, the onset of coherent oscillations can be delayed. Then, 
the axion dark matter with a small decay constant is possible. 
We also discuss the $\pi$\hspace{-0.2mm}nflation mechanism to realize the hilltop initial condition in the stochastic axion scenario.
}

\begin{document}
\begin{flushright}
      \normalsize TU--1097
\end{flushright}
\maketitle
\flushbottom

%%%%%%%%%%%%%%%%%%%%%%%%%%%%%%%%%%%%%%%%%%%%%%%%%%%%%%%%%%%%%%%%
\section{Introduction
\label{sec:introduction}}
%%%%%%%%%%%%%%%%%%%%%%%%%%%%%%%%%%%%%%%%%%%%%%%%%%%%%%%%%%%%%%%%
The identity of dark  matter remains a great mystery. 
Dark matter is known to be very stable, and its lifetime must be much longer than the present
age of the universe~\cite{Aghanim:2018eyx}. The stability of dark matter can be explained in various ways, 
and one plausible possibility is that it is due to the small mass and feeble interactions with the standard model particles. 

In the string theory, a large number of moduli or axions appear in the low-energy effective theory after compactifying
the extra dimensions. Some of them may remain so light that they play an important role in cosmology such as the inflaton,
dark energy, or dark matter. Indeed, an axion is known to be a plausible candidate for dark matter;
it is stable  on cosmological time scales due to its small mass and feeble interactions, and moreover, 
it can be copiously produced by the vacuum misalignment mechanism~\cite{Preskill:1982cy,Abbott:1982af,Dine:1982ah}. In this paper we study the string axion (simply axion
hereafter) as a dark matter candidate.

The axion abundance depends on the initial misalignment angle and the global shape of the potential. Usually one adopts the initial
angle $\theta_{\rm ini}$ of order unity to estimate the axion abundance. Recently, however, it was pointed out in Refs.~\cite{Graham:2018jyp,
Guth:2018hsa} (including 
two of the present authors F.T. and W.Y.) that the typical value of $\theta_{\rm ini}$ can be naturally much smaller than unity if the inflation scale is
low and if the inflation lasts sufficiently long. This is because the stochastic behavior of the axion is balanced by the classical dynamics
after sufficiently long inflation, and
its probability distribution reaches equilibrium, the so-called Bunch-Davies
(BD) distribution~\cite{Bunch:1978yq} peaked at the potential minimum.
In some sense, the axion knows where the minimum is in a probabilistic way. 
The BD distribution was applied to the string axion in Ref.~\cite{Ho:2019ayl}, in which it was shown that the moduli problem 
induced by the string axion can be significantly alleviated.

The other important factor of the axion abundance is the shape of the axion potential. 
If there are many axions with mass and kinetic mixings, they may form a complicated axion landscape~\cite{Higaki:2014pja,Higaki:2014mwa},
which has interesting implications for
 inflation models~\cite{Higaki:2014pja,Higaki:2014mwa,Wang:2015rel,Masoumi:2016eqo,Nath:2017ihp,Yamada:2017uzq} and dark matter~\cite{Daido:2016tsj}. In order for one of the axions
to explain dark matter, its mass must be extremely light compared to the fundamental scale. 
Broadly speaking, such a light axion mass
can be realized in the following two cases: (i) there is a flat direction in the axion landscape, along which the
axion potential is extremely flat; (ii) the axion mass is suppressed in the vicinity of a potential minimum
due to cancellation among different contributions. In the first case, the  potential  is well approximated by a
quadratic term when expanded around the minimum, and the axion abundance in this case is the one usually adopted
in the literature. In the second case, on the other hand, the light axion mass is just a consequence of cancellation,
and it implies that the quartic coupling is not generally suppressed. Therefore, it is likely that the potential can be
approximated by a quartic potential plus  a tiny mass term when expanded around the  minimum. 
The cancellation may be due to the anthropic requirement for dark matter. 
In Ref.~\cite{Daido:2016tsj} 
 the axion abundance and its isocurvature fluctuations were studied in a set-up corresponding to the second case.
 It was assumed that the initial misalignment angle was such that the curvature of the potential becomes comparable to the Hubble parameter during inflation. However, it was not studied how the viable parameter region will be modified if
 one uses the BD distribution as the initial condition. The suppressed initial angle has two effects. 
 One is to suppress the axion abundance. The other is to enhance the isocurvature perturbation for a fixed inflation scale.
 The  purpose of this paper is to study these effects in detail and show the viable parameter space in this scenario.

In this paper we study the stochastic axion scenario where the initial misalignment angle follows the probability distribution
determined by the competition between the quantum diffusion and the classical motion. We allow the axion potential
to deviate from a simple quadratic potential, and estimate the axion abundance and isocurvature perturbations to delineate the  
viable parameter space. We will  take account of various cosmological bounds such as non-detection of
the primordial gravitational waves and the upper bound on the extra diffuse X-ray/$\gamma$-ray fluxes. As we shall see, a broader
range of the inflation scale is allowed in the case where the axion potential is approximated by the quartic plus tiny quadratic terms
compared to the  conventional case with the simple mass term. 

The rest of this paper is organized as follows. In Sec.~\ref{sec:setup} we explain the set-up for the axion potential.  After a brief 
review of the stochastic axion scenario, we estimate the initial misalignment angle and the resultant axion abundance in Sec.~\ref{sec:abundance}. In Sec.~\ref{sec:bounds} various cosmological bounds are taken into account to delineate the viable parameter space. The last two sections are devoted to discussion and conclusions.

%%%%%%%%%%%%%%%%%%%%%%%%%%%%%%%%%%%%%%%%%%%%%%%%%%%%%%%%%%%%%%%%
\section{Light axion in the axion landscape
\label{sec:setup}}
%%%%%%%%%%%%%%%%%%%%%%%%%%%%%%%%%%%%%%%%%%%%%%%%%%%%%%%%%%%%%%%%
Let us suppose that there are many axions which have mass and kinetic mixings and constitute a complex landscape,
the so-called axion landscape~\cite{Higaki:2014mwa, Higaki:2014pja}. 
The axion potential can be modeled by
\begin{eqnarray}
V(\phi_\alpha)=\sum_{i=1}^{N_{\rm{s}}}\Lambda^4_i\left(1-\cos\left(\sum_{\alpha=1}^{N_{\rm{A}}}n_{i\alpha}\frac{\phi_\alpha}{f_\alpha}+\delta_i\right)\right)+C,
\end{eqnarray}
where the prefactor $\Lambda_i$ and  $\delta_i$ respectively represent the dynamical scale and a CP phase of the corresponding non-perturbative effect, $n_{i\alpha}$ is an integer-valued anomaly coefficient matrix, $f_\alpha$ is the decay constant, and $N_{\rm{s}}$ and $N_{\rm{A}}$ are the number of shift symmetry breaking terms and axions, respectively. The constant term $C$ is chosen so that  the cosmological constant is vanishingly small in the present universe. 
In this paper we assume that one of the axions is responsible for the observed dark matter.
To this end, the axion must be sufficiently long-lived, implying that its mass is very light.
Broadly speaking, there are two possibilities to realize such a light axion in the axion landscape.
 One possibility is that there exists a flat direction in the landscape. It may be that the corresponding 
 dynamical scale happens  to be much smaller than the other directions, or the effective decay constant may be
 enhanced by the KNP mechanism~\cite{Kim:2004rp,Choi:2014rja,Higaki:2014mwa}.
Although the enhanced decay constant certainly  makes the axion lighter, we consider the former case because the axion mass must be extremely light to have a lifetime longer than the
present age of the universe, and the latter would require a rather contrived set-up to realize such a
large hierarchy in the effective decay constants. 
Focusing on the flat direction,  we consider the potential modeled by
a single cosine term,
\beq
V(\phi)=\Lambda^4\left[1-\cos\left(\frac{\phi}{f_{\phi}}\right)\right]\simeq\frac{1}{2}m_{\phi}^2\phi^2,
\label{single}
\eeq
where $\Lambda$ denotes the potential height, $f_\phi$ is  the axion decay constant, and 
we have expanded the potential at the origin in  the second equality for 
$|\phi| \lesssim f_{\phi}$. The axion mass $m_\phi$ is given  by
 \beq
 m_{\phi} = \frac{\Lambda^2}{f_\phi},
 \eeq
and we assume $\Lambda \ll f_\phi$.
 
The other possibility is that the axion mass at one of the potential minima in the landscape happens to be
much smaller than the typical curvature scale.
This can be realized by cancellation among several shift symmetry breaking terms. For example, a single axion with two shift symmetry breakings satisfying this property was considered in Ref.~\cite{Daido:2016tsj}, and its potential is given by
\begin{eqnarray}
V(\phi)=\Lambda_1^4\left[1-\cos\left(n_1\frac{\phi}{f_\phi}\right)\right]+\Lambda_2^4\left[1-\cos\left(n_2\frac{\phi}{f_\phi}+\delta\right)\right]+{\rm const.},
\label{potential1}
\end{eqnarray}
where $\Lambda_1$ and $\Lambda_2$ represent the size of the shift symmetry breakings, 
and $\delta$ is the CP phase between the two terms.
The axion mass can be suppressed if the contributions of the two terms
are almost canceled at the minimum.
Specifically we consider $\delta=\pi$ and $n_2>n_1$, and expand the potential  around the minimum at 
$\phi = 0$ as
\begin{eqnarray}
V(\phi)\simeq\frac{1}{2}m_{\phi}^2\phi^2+\frac{\lambda}{4!}\phi^4
\label{pot1app}
\end{eqnarray}
for $|\phi| < f_\phi$.
Here, the axion mass $m_\phi$ and the quartic coupling $\lambda$ are respectively given by
\begin{align}
m_{\phi}^2 &= \frac{n_1^2\Lambda_1^4-n_2^2\Lambda_2^4}{f_\phi^2},\\
\lambda &=\frac{n_2^4\Lambda^4_2-n_1^4\Lambda^4_1}{f_\phi^4}\simeq(n_2^2-n_1^2)\frac{n_1^2\Lambda_1^4}{f_\phi^4},
\label{lambda}
\end{align}
where in the last equality, we have used the fact that the two shift symmetry breaking terms 
are almost canceled to realize the light axion mass, i.e.,  
$$\frac{n_1^2\Lambda_1^4}{f_\phi^2}\simeq\frac{n_2^2\Lambda_2^4}{f_\phi^2}\gg m_{\phi}^2.$$
On the other hand, the quartic coupling and higher order terms are not suppressed in general.
To be concrete, we will set $n_1=1$ and $n_2=2$ in which case the periodicity of the potential is 
equal to $2\pi f_\phi$.

For later use we also define a dimensionless angle, 
$$\theta \equiv \frac{\phi}{f_\phi}.$$
The axion potential changes from the quadratic to quartic term around $\theta \simeq \theta_{\rm{tr}}$ defined by
\beq
\theta_{\rm{tr}}\equiv\frac{m_{\phi}}{f_{\phi}}\sqrt{\frac{6}{\lambda}}\simeq\sqrt{\frac{6}{n_2^2-n_1^2}}\frac{f_{\phi}m_{\phi}}{\Lambda_1^2}\ll 1.
\label{transition}
\eeq
Precisely speaking, we define $\theta_{\rm tr}$ such that 
the derivative of the quadratic term becomes equal to that of the quartic term at $\theta = \theta_{\rm tr}$,
 since we are interested in the transition 
of the axion oscillations whose periodicity depends on $V'(\phi)$ at the oscillation amplitude.

%%%%%%%%%%%%%%%%%%%%%%%%%%%%%%%%%%%%%%%%%%%%%%%%%%%%%%%%%%%%%%%%
\section{Bunch-Davies distribution and axion abundance
\label{sec:abundance}}
%%%%%%%%%%%%%%%%%%%%%%%%%%%%%%%%%%%%%%%%%%%%%%%%%%%%%%%%%%%%%%%%
If the axion is light during inflation, it
acquires quantum fluctuations of order the Hubble parameter, $H_{\rm inf}$. The fluctuations 
continuously exit the horizon and become classical soon afterwards.
Those fluctuations can be treated as a random Gaussian noise.
Therefore, while the axion is classically driven  toward the potential minimum, it is also randomly kicked upward or downward on the potential by the small-scale fluctuations. 
Such stochastic axion dynamics can be well described by the Fokker-Planck equation that takes account of
the effect of the random Gaussian noise as a diffusion effect on the axion probability distribution.  
Then, the two competing effects end up with an equilibrium distribution of the axion field, if inflation lasts long enough. This probability distribution is called the BD distribution~\cite{Bunch:1978yq}. It was shown 
in Refs.~\cite{Graham:2018jyp,Guth:2018hsa} that the QCD axion abundance can be suppressed if the initial angle follows
the BD distribution. 

In the following we derive the BD distribution of the initial angle $\theta_{\rm{ini}}$ 
as a function of the inflation scale for a general axion potential. Then, we estimate the axion abundance
by applying it  to  the quadratic or quartic potentials.

\subsection{Bunch-Davies distribution
\label{sec:BDdist}}
We assume that the Hubble parameter during inflation, $H_{\rm inf}$, is approximately constant in time, and derive the BD  distribution for the axion field, $\phi$. First, let us separate it into the long and short wave-length modes, $\phi=\bar{\phi}+\delta\phi_{\rm{short}}$. The axion dynamics under the effect of short-wavelength fluctuations is described by the Langevin equation,
\begin{eqnarray}
\dot{\bar{\phi}}=-\frac{1}{3H_{\rm{inf}}}V'(\bar{\phi})+f(\bm{x} , t),
\label{Langevin}
\end{eqnarray}
where $V(\phi)$ is a periodic potential of $\phi$ and the dot and prime represent the derivative with respect to the cosmic time $t$ and axion field $\phi$, respectively.
We assume that $V(\phi)$ is negligibly small compared to the total energy density of the universe,
and it  satisfies  $|V''(\phi)|\ll H^2_{\rm{inf}}$
for all values of $\phi$ so that the stochastic formalism is applicable.
The information of the short wave-length mode $\delta\phi_{\rm{short}}$ is included in the Gaussian noise term, $f(\bm{x},t)$,
satisfying
\begin{eqnarray}
\left\langle f(\bm{x} , t_1)f(\bm{x} , t_2)\right\rangle=\frac{H_{\rm{inf}}^3}{4\pi^2}\delta(t_1-t_2),
\end{eqnarray}
where $\langle \cdots \rangle$ represents the stochastic average. 
The corresponding Fokker-Planck equation is given by
\begin{eqnarray}
\frac{\partial\mathcal{P}(\phi, t)}{\partial t}=\frac{1}{3H_{\rm{inf}}}\frac{\partial}{\partial\phi}(V'(\phi)\mathcal{P}(\phi, t))+\frac{H_{\rm{inf}}^3}{8\pi^2}\frac{\partial^2\mathcal{P}(\phi, t)}{\partial\phi^2},
\end{eqnarray}
where $\mathcal{P}(\phi, t)$ denotes the probability distribution for the coarse-grained field $\phi$. 

We are interested in the asymptotic form of the probability distribution, $\mathcal{P}_{\rm BD}(\phi)$, which satisfies
\begin{eqnarray}
\frac{1}{3H_{\rm{inf}}}V'(\phi)\mathcal{P}_{\rm BD}(\phi)+\frac{H_{\rm{inf}}^3}{8\pi^2}\frac{\partial\mathcal{P}_{\rm BD}(\phi)}{\partial\phi} = 0.
\end{eqnarray}
This equation follows from the Fokker-Planck equation with $\partial\mathcal{P}(\phi, t)/\partial t=0$ 
 under  the assumption that
$V(\phi)$ and $\mathcal{P}_{\rm BD}(\phi)$ are periodic with respect to $\phi$.\footnote{Alternatively, one may assume that the probability distribution vanishes at the boundary of $\phi$. }
The solution is given by
\begin{eqnarray}
\label{BDfull}
\mathcal{P}_{\rm BD}(\phi) \propto \exp\left( - \frac{8 \pi^2}{3 H_{\rm inf}^4} V(\phi) \right).
\end{eqnarray}
This is called the BD distribution.
Therefore, the BD distribution is the largest where $V(\phi)$ is the smallest.

For convenience, we define the probability distribution for the dimensionless 
angle $\theta = \phi/f_\phi$ as
\begin{align}
P(\theta) \propto \mathcal{P}_{\rm BD}(\theta f_\phi).
\end{align}
In the following we set the origin of $\theta$ equal to one of the minima of $V(\phi)$,
and consider the probability 
distribution around it in the range of $\theta_- \leq \theta \leq \theta_+$. 
The  boundary is given by $\theta_\pm = \pm \pi$ for the two potentials given in the previous section.
The proportionality factor is determined by
the normalization condition,
\begin{align}
\int_{\theta_{-}}^{\theta_{+}} P(\theta) d\theta = 1,
\end{align}
where we implicitly assume that $P(\theta)$ is vanishingly small at the boundary.
We will comment on the case where this is not satisfied at the  end of this subsection.
Using the above probability distribution,  we can define the typical initial angle as
the variance,
\begin{eqnarray}
\theta_{\rm{ini}}\equiv \sqrt{\langle\theta^2\rangle}= \left(\int_{\theta_{-}}^{\theta_{+}} \theta^2P(\theta)d\theta
\right)^{1/2}.
\label{inidef}
\end{eqnarray}

In the case of the quadratic potential (\ref{single}), the probability distribution is given  by the Gaussian form,
\begin{eqnarray}
P(\theta)\propto
\exp\left[\displaystyle{-\frac{4\pi^2m_{\phi}^2f_{\phi}^2}{3H_{\rm{inf}}^4}\theta^2}\right]~~~~~(|\theta| \lesssim 1),
\label{distribution1}
\end{eqnarray}
where one can see that the distribution is peaked at the potential minimum $\theta=0$, and it is  unlikely
for $\theta$ to take values much greater than $H_{\rm inf}^2/m_\phi f_\phi$. Using (\ref{inidef}) and (\ref{distribution1}), one can estimate the initial angle $\theta_{\rm{ini}}$ as
\beq
\theta_{\rm{ini}}\simeq 1.9\times10^{-6} \left(\frac{H_{\rm{inf}}}{10^5\GeV}\right)^2\left(\frac{f_{\phi}}{10^{16}\GeV}\right)^{-1}\left(\frac{m_{\phi}}{100\MeV}\right)^{-1},
\label{thetainiquad}
\eeq
which can be naturally much smaller than unity without fine-tuning the initial condition. Note that this expression is valid
only for $\theta_{\rm ini} \lesssim 1$. 

Similarly, in the case of the potential given by (\ref{pot1app}), 
the probability distribution is approximately given  by
\begin{eqnarray}
P(\theta)\propto
\begin{cases}
\exp\left[\displaystyle{-\frac{\lambda\pi^2f_{\phi}^4}{9H_{\rm{inf}}^4}\theta^4}\right]
\hspace{1cm} &(\theta_{\rm{tr}} \lesssim |\theta| \lesssim 1)
\\
&\\
\exp\left[\displaystyle{-\frac{4\pi^2m_{\phi}^2f_{\phi}^2}{3H_{\rm{inf}}^4}\theta^2}\right]
&(|\theta| \lesssim \theta_{\rm{tr}})
\end{cases}.
\label{distribution}
\end{eqnarray}
Using (\ref{inidef}) and (\ref{distribution}), one can estimate the initial angle $\theta_{\rm{ini}}$ as
\begin{empheq}[left={\theta_{\rm{ini}}\simeq\empheqlbrace}]{align}
&4.3\times10^{-5}\left(\frac{H_{\rm{inf}}}{10^8\GeV}\right)\left(\frac{\Lambda_1}{10^{12}\GeV}\right)^{-1}
&(\theta_{\rm{tr}}\lesssim \theta_{\rm{ini}}\lesssim 1)\label{initial1}\\
&1.9\times10^{-6} \left(\frac{H_{\rm{inf}}}{10^5\GeV}\right)^2\left(\frac{f_{\phi}}{10^{16}\GeV}\right)^{-1}\left(\frac{m_{\phi}}{100\MeV}\right)^{-1}
&(\theta_{\rm{ini}}\lesssim  \theta_{\rm{tr}}).
\label{initial2}
\end{empheq}

\begin{figure}[t!]
\includegraphics[width=10cm]{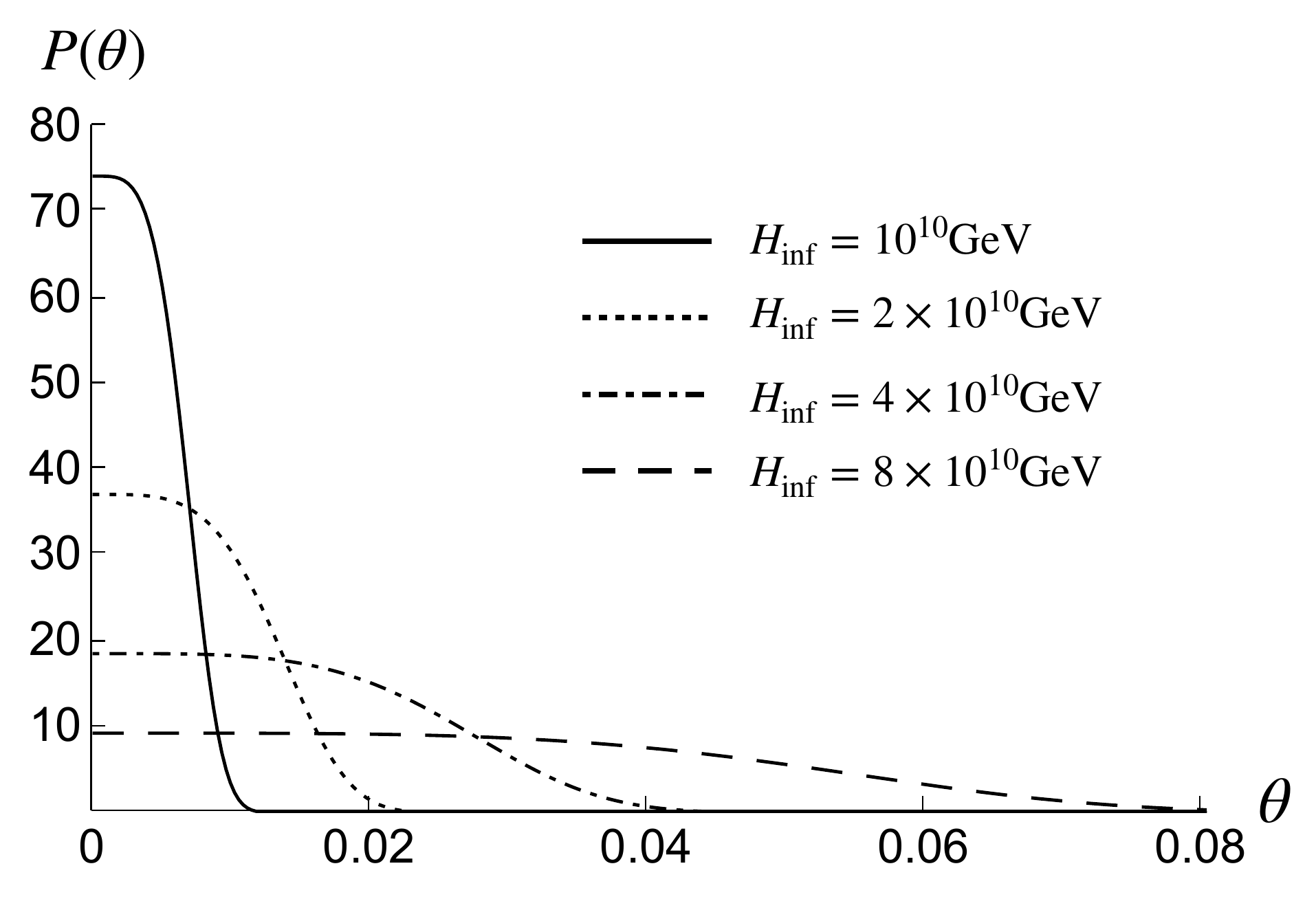}
\centering
\caption{The BD distribution as a function of $\theta$ for several values of $H_{\rm{inf}}$ 
in the case of the approximated potential (\ref{pot1app}). We set $\lambda=3 \times 10^{-16}$,
$f_\phi=10^{16}\GeV$, and $m_\phi=1\eV$,  corresponding to $\theta_{\rm{tr}} \simeq 1.4 \times 10^{-17}$.
}
\label{dist}
\end{figure}

We show the probability distribution $P(\theta)$ in Fig.~\ref{dist} for several values of $H_{\rm inf}$. This is the case that the axion populates the approximated potential (\ref{pot1app}) with $f_\phi=10^{16}\GeV$, $\lambda=3\times10^{-16}$, and $m_\phi=1\eV$. 
One can see that the distribution gets broader as $H_{\rm{inf}}$ increases. This is because the
diffusion effect is stronger for larger $H_{\rm inf}$. In other words, the initial angle is sharply peaked
at the potential minimum for a sufficiently small $H_{\rm{inf}}$.

\begin{figure}[t!]
\includegraphics[width=12cm]{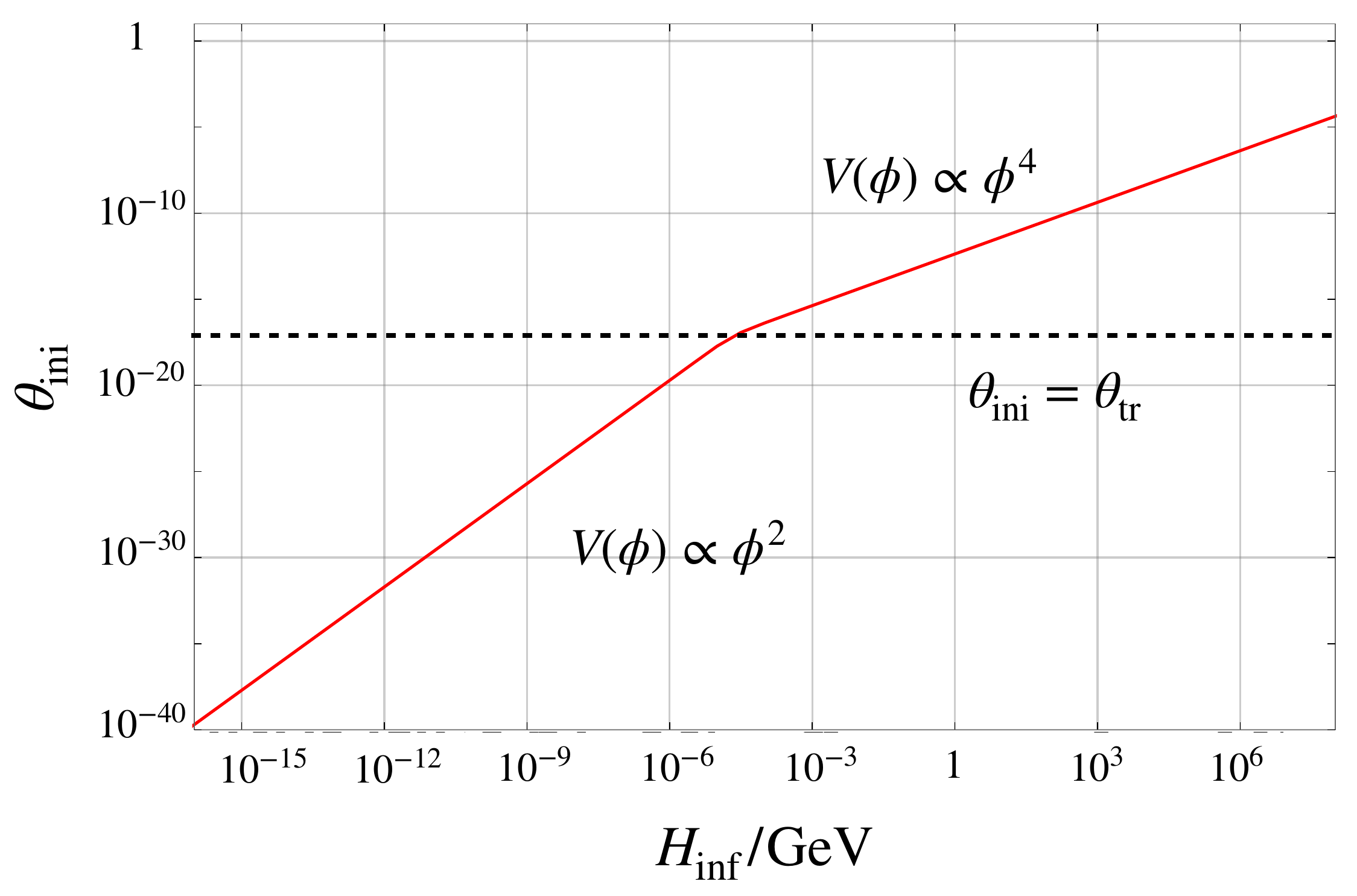}
\centering
\caption{The initial angle $\theta_{\rm{ini}}$ as a function of the inflation scale $H_{\rm{inf}}$ for $\Lambda_1=10^{12}\GeV$, $f_{\phi}=10^{16}\GeV$, and $m_{\phi}=1\eV$. The red line represents the initial angle which is estimated with the original potential (\ref{potential1}). The horizontal dotted line is the transition point, $\theta_{\rm{tr}}\simeq 1.4\times10^{-17}$.}
\label{ini1}
\end{figure}

In Fig.~\ref{ini1} we show $\theta_{\rm ini}$ as a function of $H_{\rm inf}$ 
for $\Lambda_1=10^{12}\GeV$, $m_{\phi}=1\eV$, and $f_{\phi}=10^{16}\GeV$.
Here we substitute the original potential (\ref{potential1}) into (\ref{inidef}) to calculate $\theta_{\rm ini}$
numerically. 
The  dotted horizontal line shows the transition point from quadratic to quartic potential,  
 $\theta_{\rm{ini}} = \theta_{\rm{tr}}$. One can see that $\theta_{\rm ini}$ tends to be suppressed when the
 potential is dominated by the quartic term  compared to the quadratic term. This can be understood by 
 noting that the potential gets steeper as the quartic term dominates over the quadratic one.

In order to reach the BD distribution, we need a sufficiently large number of $e$-folds;
$N\sim H_{\rm{inf}}^2/m_\phi^2$ for the quadratic potential, and $N\sim1/\sqrt{\lambda}$
for the quartic potential. 
Such a large number of $e$-folds can be realized by 
the eternal inflation~\cite{Linde:1982ur,Steinhardt:1982kg,Vilenkin:1983xq,Linde:1986fc,Linde:1986fd,Goncharov:1987ir} 
(see also \cite{Guth:2000ka,Guth:2007ng,Linde:2015edk}). Note that the eternity of the
eternal inflation usually relies on the volume measure. Recently it was shown that 
a sufficiently large number of $e$-folds can be realized without relying on the
 volume measure, if one considers
 a hilltop-type stochastic inflation with a shallow local minimum around the potential maximum~\cite{Kitajima:2019ibn}.

Lastly let us comment on the assumption that 
the probability distribution $P(\theta)$ is vanishingly small 
at $\theta = \pm \pi$ in the above discussion.
The approximated expressions \eqref{thetainiquad},  \eqref{initial1}, and \eqref{initial2} 
can only be applied to the case of $\theta_{\rm ini} \ll \pi$, or equivalently, $H_{\rm{inf}} \lesssim \Lambda$ and $\Lambda_{1,2}$.
 For $H_{\rm{inf}} > \Lambda$ or $\Lambda_{1,2}$, the axion can go over the potential barrier due to the large fluacutations and 
 all the physically inequivalent vacua will be populated as a result of the stochastic dynamics. For the potentials considered here,
the asymptotic distribution will be almost uniform over the entire values of $\phi$, and a typical value
of the initial misalignment angle measured from the nearest minimum will be of order unity. 
Also, there is thermal radiation  with the Gibbons-Hawking temperature $T_{\rm{inf}}={H_{\rm{inf}}}/{2\pi}$~\cite{Gibbons:1977mu}, which may significantly modify
the axion potential if  $\Lambda$ or $\Lambda_{1,2}$ corresponds to the dynamical scale of the non-perturbative effects responsible for generating the axion potential. In the following,  therefore, we will
focus on the case of $\theta_{\rm ini} \lesssim 1$, or equivalently,  $H_{\rm{inf}} \lesssim \Lambda_1$ or $\Lambda$.

\subsection{Axion abundance
\label{subsec:abundance}}
During inflation the axion follows the BD distribution derived in the previous subsection. After inflation, the axion
starts to oscillate about the potential minimum when the effective mass, $|V''(\phi)|$, becomes comparable
to the Hubble parameter.\footnote{\label{ftnt}
The commencement of oscillations is delayed if the initial position is
close to the potential maximum, in which case the isocurvature perturbation as well as its non-Gaussianity 
are enhanced~\cite{Kobayashi:2013nva}. See also  discussion in Sec.~\ref{sec:5}.} In the case of the axion potential (\ref{single}), the axion coherent
oscillations behave as cold dark matter. On the other hand, in the case of the axion potential  (\ref{pot1app}),
the axion coherent oscillations behave as radiation until the oscillation amplitude becomes so small  that 
the tiny mass term comes to dominate the potential.

The final axion abundance depends on the oscillation amplitude, $\phi_{\rm osc}$,  when the axion starts to oscillate.
Here let us relate $\theta_{\rm osc} \equiv \phi_{\rm osc}/f_\phi$ to the initial angle $\theta_{\rm ini}$, following 
Refs.~\cite{Daido:2016tsj, Kawasaki:2011pd}. We define the onset of the oscillation to be the time when the temporal variation of the axion field over
the Hubble time  becomes comparable to the distance to the potential minimum:
 \begin{eqnarray}
\left|H^{-1}\frac{\dot{\phi}}{\phi}\right|_{\rm{osc}}\equiv\frac{1}{2}.
\label{oscdef}
\end{eqnarray}
The choice of $1/2$ in the right-handed side is just a convention, and it can be replaced with a constant of order unity.

Before the onset of the oscillation, the axion dynamics is described by 
an attractor solution of
\begin{eqnarray}
cH\dot{\phi}\simeq -V',
\label{atteq}
\end{eqnarray}
where $c$ is a numerical coefficient that depends on the equation of state of 
the dominant component of the universe,
\begin{eqnarray}
c=
\begin{cases}
3~~~(\rm{de ~Sitter})&\\
\frac{9}{2}~~~(\rm{matter ~dominant})&\\
5~~~(\rm{radiation ~dominant})&
\end{cases}.
\end{eqnarray}
Here the equation of motion is valid if  $|V''| \ll H^2$. 
Using (\ref{oscdef}) and (\ref{atteq}) we can express the Hubble parameter at the onset of oscillations in terms
of $\phi_{\rm osc}$,
\begin{eqnarray}
H^2_{\rm{osc}}\simeq\frac{2V'(\phi_{\rm{osc}})}{c_{\rm osc} \phi_{\rm{osc}}},
\label{hosc}
\end{eqnarray}
where $c_{\rm osc}$ represents $c$  evaluated at the onset of oscillations.
For the quadratic and quartic potentials, it is given by
\begin{empheq}[left={H_{\rm{osc}}\simeq  \empheqlbrace}]{align}
& \sqrt{\frac{2}{c_{\rm osc}}} m_\phi
&{\rm (quadratic)}\label{oscini1}\\
&\sqrt{\frac{\lambda}{3 c_{\rm osc}}} \phi_{\rm osc}
&{\rm (quartic)}.
\label{hosc2}
\end{empheq}
Integrating Eq.~(\ref{atteq}) from the end of inflation 
to the onset of oscillations, we obtain
\begin{eqnarray}
\int_{\phi_{\rm{ini}}}^{\phi_{\rm{osc}}}\frac{d\phi}{V'}\simeq-\frac{1}{2c_{\rm osc} (c_{\rm osc}-3)H_{\rm{osc}}^2},
\label{int}
\end{eqnarray}
where we have used $\dot{H}/H^2=3-c$, and $H_{\rm osc} \ll H_{\rm inf}$.
Then we can relate $\theta_{\rm{osc}}$ to $\theta_{\rm ini}$ for the quadratic 
and quartic potentials,
\begin{empheq}[left={\theta_{\rm{osc}}\simeq \theta_{\rm{ini}} \times \empheqlbrace}]{align}
& \exp\left[-\frac{1}{4(c_{\rm osc}-3)}\right]
&{\rm (quadratic)}\label{oscini1}\\
&\sqrt{\frac{2c_{\rm osc}-7}{2c_{\rm osc}-6}}
&{\rm (quartic)},
\label{oscini2}
\end{empheq}
One can see that $\theta_{\rm osc} \simeq \theta_{\rm ini}$ in these cases. 
Since $\theta_{\rm ini}$ is determined by the BD distribution in our scenario, the final axion abundance
depends on the inflation scale, $H_{\rm inf}$. Note that, in the previous work \cite{Daido:2016tsj},
 the initial angle $\theta_{\rm ini}$ was chosen in such a way that the curvature of the axion potential 
 at $\theta=\theta_{\rm ini}$ becomes comparable
 to the Hubble parameter during inflation. On the other hand,  $\theta_{\rm ini}$ is dynamically determined 
 by the competition between the classical motion and quantum diffusion in our case. As a result, the initial
 angle is  much smaller than assumed in the Ref.~\cite{Daido:2016tsj}.
In the following we estimate the axion abundance
for the quadratic and quartic potentials using the oscillation amplitude determined by the BD distribution.

\subsubsection{The case of the quadratic potential}
First, let us consider the case in which the axion potential is well approximated by the quadratic potential. This is the case 
either if the axion potential is given by a single cosine term (\ref{single}) and the initial angle satisfies $\theta_{\rm ini} 
\lesssim 1$, or if the potential is given by a flat-bottomed form (\ref{pot1app}) and $\theta_{\rm ini}$ is
smaller than $\theta_{\rm tr}$.

The axion abundance depends on whether the reheating is completed or not at the onset of oscillations.
First, let us consider the case in which the axion starts to oscillate before the reheating. 
Taking account of the conservation of the entropy in the comoving volume after reheating, and the dependence of the axion energy density on the scale factor $R$, $\rho_\phi\propto R^{-3}$, the ratio of the axion energy density to the entropy density  at present is  given by 
\beq
\left.\frac{\rho_{\phi}}{s}\right|_0=
\left.\frac{\rho_{\phi}}{s}\right|_{\rm reh}=
\left.\frac{\rho_{\rm inf}}{s}\right|_{\rm reh}
\left.\frac{\rho_{\phi}}{\rho_{\rm{inf}}}\right|_{\rm osc} \simeq
 \frac{3 T_{\rm reh}}{4} \frac{\rho_{\phi, \rm{osc}}}{3 H_{\rm{osc}}^2 \Mpl^2},
\eeq
where the subscripts, `0', `reh', and `osc', imply that the variables are evaluated at present, reheating, and the onset of oscillations, respectively,
and $\rho_{\rm inf}$ is the inflaton energy density. For simplicity we have assumed that the inflaton energy density decreases
as non-relativistic matter before the reheating, and that the reheating takes place almost instantaneously at $T=T_{\rm{reh}}$. 
The axion energy density at the onset of oscillations is given by
\beq
\rho_{\phi, \rm{osc}}&=&\frac{1}{2}\dot{\phi}_{\rm{osc}}^2+\frac{1}{2}m_{\phi}^2\phi_{\rm{osc}}^2=\left(\frac{1}{4c}+\frac{1}{2}\right)m_\phi^2\phi^2_{\rm{osc}}.
\label{phiosc}
\eeq 
Using (\ref{thetainiquad}), (\ref{hosc}), and (\ref{phiosc}), we can express  the axion abundance in terms of the
density parameter as
\beq
\Omega_\phi h^2
\simeq 0.39 \,\left(\frac{H_{\rm{inf}}}{10^6\GeV}\right)^4\left(\frac{m_{\phi}}{1\GeV}\right)^{-2}\left(\frac{T_{\rm{reh}}}{10^{6}\GeV}\right).
\label{abundance1}
\eeq
Here the density parameter is defined by $\Omega_\phi \equiv \rho_{\phi,0}/\rho_{\rm{crit}}$, 
where $\rho_{\rm crit} \simeq (0.0300 {\rm \,eV})^4 h^2$ is the critical density, and $h$ is the reduced Hubble
constant.

Next, we consider the case where the axion starts to oscillate after the reheating. 
The entropy density at the onset of the oscillations $s_{\rm{osc}}$ is given by $s_{\rm{osc}}=({2\pi^2}/{45}) g_{\rm{*}}(T_{\rm{osc}})T_{\rm{osc}}^3$ in terms of  the temperature $T_{\rm{osc}}$,
\begin{align}
T_{\rm{osc}} &\equiv \left(\frac{\pi^2g_*(T_{\rm{osc}})}{90}\right)^{-\frac{1}{4}}\sqrt{H_{\rm osc} \Mpl}, \\
&\simeq 6.7 \times 10^8 {\rm \,GeV} \lrfp{g_*(T_{\rm osc})}{106.75}{-\frac{1}{4}} \lrfp{m_\phi}{1\,{\rm GeV}}{\frac{1}{2}},
\end{align}
where $g_{\rm{*}}$ counts the effective relativistic degrees of freedom,
and we substituted $c_{\rm osc} = 5$ in the second equality. The axion abundance is given  by
\beq
\Omega_\phi h^2\simeq 0.19  \left(\frac{g_*(T_{\rm osc})}{106.75}\right)^{-\frac{1}{4}}
\left(\frac{H_{\rm{inf}}}{10^5\GeV}\right)^4\left(\frac{m_\phi}{0.3 \GeV}\right)^{-\frac{3}{2}}.
\label{abundance2}
\eeq
Note that, although it is not explicitly shown in the final results,  
$\Omega_\phi$ is proportional to $\theta_{\rm ini}^2$ in both cases.
(The dependence on $\theta_{\rm ini}$ can be inferred by noting $\theta_{\rm ini} \propto H_{\rm inf}^2$
in the case of the quadratic potential.)

\subsubsection{The case of the quartic potential}
Now we consider the flat-bottomed potential where the potential (\ref{pot1app}) consists of the quartic 
coupling plus a suppressed mass term. We focus on the case of $\theta_{\rm ini} \simeq  \theta_{\rm osc} \gtrsim \theta_{\rm tr}$ because
otherwise the axion abundance is reduced to the case of the quadratic potential.

First, let us consider the case in which the axion starts to oscillate before the reheating. 
Noting that the axion energy density decreases as $\rho_\phi\propto R^{-4}$ until the quadratic term comes to
dominate the potential, one can express the ratio of the axion energy density to the entropy density at present as
\begin{eqnarray}
\left.\frac{\rho_{\phi}}{s}\right|_{0}
= \rho_{\phi,{\rm tr}}^{1/4} \left.\frac{\rho_{\rm inf}}{s}\right|_{\rm reh}
\left.\frac{\rho_{\phi}{}^{3/4}}{\rho_{\rm{inf}}}\right|_{\rm osc}.
\label{trlreh}
\end{eqnarray}
Here $\rho_{\phi, \rm{tr}}$ is the axion energy density when the oscillation amplitude becomes equal to $\phi_{\rm tr} = \theta_{\rm tr} f_\phi$, and it is given by
\beq
\rho_{\phi,\rm{tr}}&=&\frac{9m_{\phi}^4}{2\lambda}.
\eeq
The axion energy density at the onset of oscillations is 
\beq
\rho_{\phi,\rm{osc}}&\simeq&\frac{1}{2}\dot{\phi}^2_{\rm{osc}}+\frac{\lambda}{4!}\phi_{\rm{osc}}^4 = 
\frac{\lambda}{4!} \left(1+\frac{1}{c}\right)\phi_{\rm{osc}}^4.
\eeq
Thus, we arrive at the axion abundance,
\beq
\Omega_\phi h^2 \simeq0.50&&\hspace{-5mm}\left(\frac{H_{\rm{inf}}}{10^8\GeV}\right)\left(\frac{m_{\phi}}{100\MeV}\right)\left(\frac{f_{\phi}}{10^{16}\GeV}\right)^{3}
\nonumber\\&\times&
\left(\frac{\Lambda_1}{10^{12}\GeV}\right)^{-3}\left(\frac{T_{\rm{reh}}}{10^{10}\GeV}\right).
\label{abundance3}
\eeq
One can see that $\Omega_\phi$ is proportional to $\theta_{\rm ini}$.
Note that the axion abundance does not depend on the order of the reheating and the transition from the quartic to the quadratic
potential.

Next, we consider the case in which the reheating precedes the onset of the axion oscillations. 
As before, the ratio of the axion energy density to the entropy density at present is given by
\begin{eqnarray}
\left.\frac{\rho_{\phi}}{s}\right|_0=\left.\frac{\rho_{\phi}}{s}\right|_{\rm tr}=\rho_{\phi,\rm{tr}}^{\frac{1}{4}}\frac{\rho^\frac{3}{4}_{\phi,\rm{osc}}}{s_{\rm{osc}}}.
\end{eqnarray}
where the temperature at the oscillation reads
\beq
T_{\rm{osc}} \simeq
3.4 \times 10^{10} {\rm\, GeV}  \lrfp{g_*(T_{\rm osc})}{106.75}{-\frac{1}{4}}
\lrfp{H_{\rm inf}}{10^8 {\rm\,GeV}}{\frac{1}{2}}
\lrfp{f_\phi}{10^{16}{\rm\,GeV}}{-\frac{1}{2}}
\lrfp{\Lambda}{10^{12}{\rm\,GeV}}{\frac{1}{2}}.
\eeq
Therefore, the axion abundance is obtained as
\begin{eqnarray}
\Omega_\phi h^2\simeq 2.0&&\hspace{-5mm}
\left(\frac{g_*(T_{\rm{osc}})}{106.75}\right)^{-\frac{1}{4}}\left(\frac{H_{\rm{inf}}}{10^8\GeV}\right)^{\frac{3}{2}}\left(\frac{m_{\phi}}{100\MeV}\right)
\nonumber\\ &\times&
\left(\frac{f_{\phi}}{10^{16}\GeV}\right)^{\frac{5}{2}} \left(\frac{\Lambda_1}{10^{12}\GeV}\right)^{-\frac{5}{2}}.
\label{abundance4}
\end{eqnarray}
One can see that $\Omega_\phi$ in this case is proportional to $\theta_{\rm ini}^\frac{3}{2}$.

Note that the dependence of the axion abundance on the inflation scale is milder than the case of the quadratic potential
(see Fig.~\ref{ini1}). This as well as the tiny $m_\phi$ are the reason why the larger $H_{\rm inf}$ can be consistent with the dark matter abundance
in the case of the quartic potential.

\subsubsection{Thermal axion particle production}
So far, we have considered the axion condensate. 
When the reheating temperature is sufficiently high, thermal axion particle production becomes important. 
If too many axion particles are produced, they contribute to hot or warm dark matter. 
Here let us estimate the abundance of thermally produced axions. 
To this end, suppose for simplicity that the axion couples to weak gauge bosons as 
\beq
{\cal L}  = c_2 \frac{\alpha_2}{8 \pi} \frac{\phi}{f_\phi}W^a_{\mu \nu} {\tilde W}_a^{\mu \nu}+ c_Y \frac{\alpha_Y}{4 \pi} \frac{\phi}{f_\phi} B_{\mu \nu} \tilde B^{\mu \nu},
\eeq
where $(\alpha_2, W^a_{\mu \nu})$ and $(\alpha_Y, B_{\mu \nu})$ are the fine-structure constant and  gauge field strength of SU(2)$_L$ and U(1)$_Y$, respectively, and  $c_2$ and $c_Y$ are model-dependent anomaly 
coefficients.\footnote{In the low energy effective theory in the broken phase, 
one obtains the coupling to photons, $\alpha_{\phi\gamma\gamma}$ defined in Eq.\,\eqref{Lagrangian} as
$ \alpha_{\phi\gamma\gamma}/\alpha_{\rm EM} = c_2/2 + c_Y$.}
Since they are higher dimensional terms, thermal production is efficient around the highest temperature, i.e. $T\sim T_{\rm reh}$, in the radiation dominant era. Here we assume the instantaneous reheating  for simplicity. 
The axion number density to entropy density is estimated
\beq
\frac{n^{\rm th}_\phi}{s}\simeq r \frac{45 \zeta{(3)}}{2 g_{*} \pi^4}
\eeq
where $r$ was calculated in \cite{Salvio:2013iaa},
\beq r\approx \left(3\times 10^{-10}c_2^2+ 5\times 10^{-11}c_Y^2\right) \left(\frac{10^{16}\,\text{GeV}}{f_\phi}\right)^2  \left(\frac{T_{\rm reh}}{10^{10}\,\text{GeV}}\right).\eeq
One obtains the abundance of the thermally produced axion as
\begin{align}
\Omega_{\phi}^{\rm th} h^2 
&\simeq \left(2\times 10^{-9} c_2^2+3\times10^{-10}  c_Y^2\right)  \left( \frac{m_\phi}{10\text{\,keV}}\right)\left(\frac{10^{16}\,\text{GeV}}{f_\phi}\right)^2  \left(\frac{T_{\rm reh}}{10^{10}\,\text{GeV}}\right).
\end{align}
Thus, the thermal production is always subdominant for $c_2,c_Y={\cal O}(1)$ in the parameter region that we will consider. 
However, it could be sizable when $T_{\rm reh}\sim f_\phi$ or $c_2, c_Y\gg {\cal O}(1)$. 
Also, thermal production can be relevant if there are other interactions e.g. an axion-top quark interaction. 

Notice that the above result applies to both potentials of \eqref{single} or \eqref{pot1app} as long as
 $\Lambda_1$ or $\Lambda$ is much smaller than $f_\phi.$
On the other hand, when $\Lambda_1$ in \eqref{pot1app} is so large that the self-interaction of $\phi$ becomes important,
 the processes of $\phi\phi\leftrightarrow \phi\phi , \phi\phi \leftrightarrow \phi \phi\phi\phi$ may not be neglected. 
 These processes increase the number density, while reducing the energy per one axion particle. Therefore in this case we can obtain 
 much colder axion particles (if the processes are in equilibrium 
 the typical temperature is $T_\phi \sim r^{1/4} T$) and their abundance is larger than the above estimate.
For $\Lambda_1\gg H_{\rm inf},$ the contribution of coherent mode, \eqref{abundance4}, can be suppressed and thus 
thermally produced axions may explain the observed dark matter.\footnote{In this case, the isocurvature bound is evaded.}

%%%%%%%%%%%%%%%%%%%%%%%%%%%%%%%%%%%%%%%%%%%%%%%%%%%%%%%%%%%%%%%%
\section{Cosmological bounds
\label{sec:bounds}}
%%%%%%%%%%%%%%%%%%%%%%%%%%%%%%%%%%%%%%%%%%%%%%%%%%%%%%%%%%%%%%%%
In this section, we study cosmological constraints on the axion dark matter in our scenario. Specifically we consider
the axionic isocurvature perturbations and the diffuse X-ray/$\gamma$-ray fluxes from the axion decays. By doing so we
will be able to identify the viable parameter region where the axion explains the observed dark matter abundance.

\subsection{Isocurvature perturbations
\label{sec:isocurvature}}
Since the axion is assumed to be light during inflation, it acquires quantum fluctuations 
of $\delta \phi_k \simeq H_{\rm inf}/2\pi$ at the horizon exit, where $k$ denotes the comoving wavenumber. 
This gives rise to the almost scale-invariant isocurvature perturbation, whose
 power spectrum is given by
\begin{eqnarray}
\mathcal{P}_S=\left[\frac{\Omega_{\phi}}{\Omega_{\rm{DM}}}\frac{\partial \ln\Omega_\phi}{\partial\theta_{\rm{ini}}}\frac{\delta\phi_{k}}{f_{\phi}}\right]^2=\left[p \frac{H_{\rm inf}}{2 \pi \theta_{\rm ini} f_{\phi}}\right]^2,
\label{isocurvature}
\end{eqnarray}
where $\Omega_{\rm DM} \simeq 0.12 h^{-2}$ is the density parameter of the dark matter~\cite{Aghanim:2018eyx},
and we have assumed  $\Omega_\phi = \Omega_{\rm DM}$ in the second equality.
Here we have defined $p \equiv (\partial \ln\Omega_\phi/\partial\theta_{\rm{ini}})$, and it is given by $p= 2$  for the quadratic potential, and $p=1$ or $3/2$ for the quartic potential (cf.  Sec.~\ref{subsec:abundance}).

A mixture of isocurvature perturbations is tightly constrained by the Planck data.
The current upper bound on the scale-invariant and uncorrelated isocurvature perturbation reads~\cite{Akrami:2018odb}
\beq
\mathcal{P}_S < 8.3 \times 10^{-11}.
\label{Planck}
\eeq
In the following we derive the upper bound on the inflation scale as a function of the axion mass
for the quadratic and quartic potential.

\subsubsection{The case of the quadratic potential}
Using (\ref{thetainiquad}) and $p=2$, we obtain
\beq
\mathcal{P}_S = \frac{8 m_\phi^2}{3 H_{\rm inf}^2} 
\simeq 2.7\times10^{-10}\left(\frac{H_{\rm{inf}}}{10^5\GeV}\right)^{-2}\left(\frac{m_\phi}{1\GeV}\right)^2.
\eeq
Thus, to satisfy the isocurvature bound (\ref{Planck}), we need
\beq
H_{\rm inf} \gtrsim 1.8 \times 10^5 \,m_\phi.
\eeq

\subsubsection{The case of the quartic potential}
We focus on the case of $\theta_{\rm osc} \gtrsim \theta_{\rm tr}$ since otherwise the result will be reduced
to the previous case. Using  (\ref{initial1}), we obtain
\beq
\mathcal{P}_S \simeq 0.14\, p^2\frac{\Lambda_1^2}{f_\phi^2} 
\simeq 0.079\, p^2 \sqrt{\lambda}.
\eeq
To satisfy the isocurvature bound (\ref{Planck}), we need
\beq
\lambda\; \lesssim \; \frac{1.1 \times 10^{-18}}{p^4}.
\eeq
Thus, the quartic coupling determines the size of isocurvature perturbations. 

Since we assume $\Omega_{\phi}=\Omega_{\rm{DM}}$, we can erase $\Lambda_1$
using (\ref{abundance3}) or (\ref{abundance4}) from the expression of $\mathcal{P}_S$.
Then, we obtain
\begin{empheq}[left={\mathcal{P}_S\simeq\empheqlbrace}]{align}
&~ 3.5 \times 10^{-9}
\left(\frac{H_{\rm{inf}}}{10^{8}\,\rm{GeV}}\right)^{\frac{2}{3}}
\left(\frac{m_{\phi}}{100\MeV}\right)^{\frac{2}{3}}
\left(\frac{T_{\rm{reh}}}{10^{10}\GeV}\right)^{\frac{2}{3}}
&(T_{\rm{reh}}\lesssim T_{\rm{osc}})\\
&~ 2.9 \times 10^{-8}
 \lrfp{g_*(T_{\rm osc})}{106.75}{-\frac{1}{5}}
\left(\frac{H_{\rm{inf}}}{10^{8}\GeV}\right)^{\frac{6}{5}}\left(\frac{m_{\phi}}{100\MeV}\right)^{\frac{4}{5}}&(T_{\rm{osc}}\lesssim T_{\rm{reh}}).
\end{empheq}
To satisfy the isocurvature bound (\ref{Planck}), we need
\begin{empheq}[left=H_{\rm inf} \lesssim \empheqlbrace]{align}
&3.6 \times 10^{5} {\rm\,GeV} \left(\frac{m_{\phi}}{100\MeV}\right)^{-1}\left(\frac{T_{\rm{reh}}}{10^{10}\GeV}\right)^{-1}
&(T_{\rm{osc}}\gtrsim T_{\rm{reh}})
\label{isoconstrBD3}\\
& 7.6 \times 10^{5} {\rm\,GeV}  \lrfp{g_*(T_{\rm osc})}{106.75}{\frac{1}{6}}
\left(\frac{m_{\phi}}{100\MeV}\right)^{-\frac{2}{3}}
& (T_{\rm{osc}}\lesssim T_{\rm{reh}}).
\label{isoconstrBD4}
\end{empheq}
In contrast to the quadratic potential, the isocurvature bound results in the upper bound on the inflation scale
in the case of the quartic potential. 

\subsection{Axion decay into photons and hidden photons
\label{sec:photons}}
The axion dark matter might have feeble interactions with the standard model or hidden sector particles. 
Here, we consider the axion-(hidden) photon couplings,
\begin{eqnarray}
\frac{\alpha_{\phi\gamma\gamma}}{4\pi f_{\phi}}\phi F_{\mu\nu}\tilde{F}^{\mu\nu},
~~~\frac{\alpha_{\phi\gamma'\gamma'}}{4\pi f_{\phi}}\phi F'_{\mu\nu}\tilde{F}'^{\mu\nu},
\label{Lagrangian}
\end{eqnarray}
where $\alpha_{\phi\gamma \gamma}$ is a model-dependent coupling constant, 
$F_{\mu\nu}$ and $\tilde{F}^{\mu\nu}$ are the field strength of
photons and its dual, respectively, and those with primes are for hidden photons. 

First let us consider the axion decay into hidden photons through the above coupling. 
The decay rate is given by
\begin{eqnarray}
\Gamma_{\phi} (\phi \to \gamma'\gamma')=\frac{\alpha_{\phi\gamma'\gamma'}^2}{64\pi^3}\frac{m_{\phi}^3}{f_{\phi}^2},
\end{eqnarray}
and if this is the dominant decay channel, the axion lifetime is given by
\begin{eqnarray}
\tau_\phi(\phi \to \gamma'\gamma') \simeq
2.5\times10^{18}\,\rm{sec} \left(\frac{\alpha_{\phi\gamma'\gamma'}}{\alpha_{\rm{EM}}}\right)^{-2}
\left(\frac{\it{m_{\phi}}}{100\,\rm{MeV}}\right)^{-3}
\left(\frac{\it{f_{\phi}}}{10^{16}\,\rm{GeV}}\right)^2,
\end{eqnarray}
where $\alpha_{\rm{EM}} \simeq 1/137$ is the electromagnetic fine-structure constant. 
The observational impact of dark matter decaying into dark radiation was 
studied in detail in Refs.~\cite{Enqvist:2015ara,Enqvist:2019tsa}, and they placed a lower bound
on the lifetime,  $\tau_{\rm{DM}}\ge175\,\rm{Gyr}$~\cite{Enqvist:2019tsa}.
This can be rewritten as the upper bound on the axion mass,
\begin{eqnarray}
m_{\phi}\lesssim76\MeV\left(\frac{\alpha_{\phi\gamma'\gamma'}}{\alpha_{\rm{EM}}}\right)^{-\frac{2}{3}}\left(\frac{\it{f_{\phi}}}{10^{16}\GeV}\right)^{\frac{2}{3}}.
\label{lifetime}
\end{eqnarray}
Thus, for $f_{\phi}\simeq10^{16}\GeV$ and $\alpha_{\phi\gamma'\gamma'}\simeq\alpha_{\rm{EM}}$,
the axion mass cannot exceed $80\,\rm{MeV}$ or so. 

Next we consider the cosmological impact of the axion decay into ordinary photons. The energetic photons produced
by the axion decay contribute to the Galactic and extra-galactic diffuse X-ray/$\gamma$-ray background. 
First, let us estimate the Galactic halo contribution to the diffuse photon background~\cite{Essig:2013goa, Ho:2019ayl}. In order to derive the expected signal, we need the Galactic dark matter density profile. Here we assume the NFW dark matter profile~\cite{Navarro:1995iw, Navarro:1996gj} and use the J-factor summarized in Table I of Ref.~\cite{Essig:2013goa}. As we consider the axion-photon interaction (\ref{Lagrangian}), the produced two photons initially have the monochromatic spectrum
\begin{eqnarray}
\frac{dN_{\gamma}}{dE_{\gamma}}=2\delta\left(E_{\gamma}-\frac{m_{\phi}}{2}\right)
\label{photonspectrum},
 \end{eqnarray}
where $E_\gamma$ and $N_{\gamma}$ are the present photon energy and number, respectively. 
 Then, the differential photon flux from the axion decay in the Milky Way is written by
 \begin{eqnarray}
 \frac{d\Phi_{\rm{MW}}}{dE_{\gamma}}=\int_{l.o.s}dy\frac{\Gamma_{\phi\gamma\gamma}}{4\pi y^2}\frac{dN_\gamma}{dE_\gamma}\frac{\rho_\phi(y)}{m_{\phi}}y^2d\Omega=\frac{r_{\odot}}{4\pi}\frac{\rho_\odot}{m_{\phi}}\Gamma_{\phi\gamma\gamma}\frac{dN_\gamma}{dE_\gamma}\frac{\Omega_\phi}{\Omega_{\rm{DM}}}J_{\rm{D}},
\label{MW}
\end{eqnarray}
where $r_\odot=8.5\rm{kpc}$ is the distance between the Sun and the center of our Galaxy, $\rho_\odot= 0.3\rm{GeV/cm^3}$ is the local dark matter density, $\rho_\phi$ is the axion energy density, and we define a J-factor as
\begin{eqnarray}
J_{\rm{D}}\equiv\int_{l.o.s}\frac{dy}{r_\odot}\frac{\rho_{\rm{DM}}(y)}{\rho_{\odot}}d\Omega,
 \end{eqnarray}
where $\rho_{\rm{DM}}$ represents a dark matter density profile. Here, the integration is taken over the observed region of the sky $\Delta\Omega$ and the line-of-sight distance $y$. 

Let us next consider the extra-galactic contributions~\cite{Asaka:1997rv}. 
The photons coming from cosmological distances suffer a redshift due to the cosmic expansion. 
Also, the dark matter density in the early universe was higher than present density. Assuming the spatially
homogeneous distribution of the axion, we obtain the photon flux per a unit solid angle 
\begin{eqnarray}
\frac{d^2\Phi_{\rm{EG}}}{d\Omega dE_\gamma}
=\frac{\Gamma_{\phi\gamma\gamma}}{4\pi}\int_{t_{\rm{rec}}}^{t_0} d\tilde{t}\, n_\phi(\tilde{t})(1+z)^{-3}\frac{d \tilde{E}_\gamma}{dE_\gamma}\frac{dN_\gamma}{d\tilde{E}_\gamma},
\label{EGint}
\end{eqnarray}
where $t_{\rm{rec}}$ is the time of the recombination, $z$ is the redshift parameter 
corresponding to the time $t= \tilde{t}$, $\tilde{E}_\gamma$ is the photon energy at the production, and the axion density $n_\phi(\tilde{t})$ is defined by
\begin{eqnarray}
n_\phi({\tilde t})\equiv n_{\phi,0}(1+z)^3=\frac{\Omega_\phi\rho_{\rm{crit}}}{m_{\phi}}(1+z)^3.
\end{eqnarray}
Here, we have neglected a slight decrease of the axion density due to the decay, because the lifetime must be
much longer than the present age of the universe.
Note that we put a cut-off at recombination, since photons cannot  propagate freely until the recombination. 
The redshift parameter $z$ is related to the cosmic time $t$ by
\begin{eqnarray}
\frac{dt}{dz}=-\left[H_0(1+z)\sqrt{\Omega_m(1+z)^3+\Omega_\Lambda}\right]^{-1},
\end{eqnarray}
where $H_0$ is the Hubble constant, and $\Omega_m$ and $\Omega_\Lambda$ denote the density parameter of non-relativistic matter and the cosmological constant, respectively. Using this relation in the integration (\ref{EGint}), 
we obtain the extra-galactic photon flux per a unit solid angle,
\begin{eqnarray}
\frac{d^2\Phi_{\rm{EG}}}{d\Omega dE_\gamma}=\frac{\Omega_\phi\rho_{\rm{crit}}\Gamma_{\phi\gamma\gamma}}{2\pi m_{\phi}E_\gamma H_0}\left[\Omega_\Lambda+\Omega_m\left(\frac{m_\phi}{2E_\gamma}\right)^3\right]^{-1/2}.
\label{EG}
\end{eqnarray}
Since we can  observe photons produced after the recombination, 
the energy $E_\gamma$ is in the range of $\frac{m_{\phi}}{2}/(1+z_{\rm{rec}})<E_\gamma<\frac{m_{\phi}}{2}$,
where $z_{\rm{rec}}\simeq1100$.

\begin{figure}[t!]
\includegraphics[width=12cm]{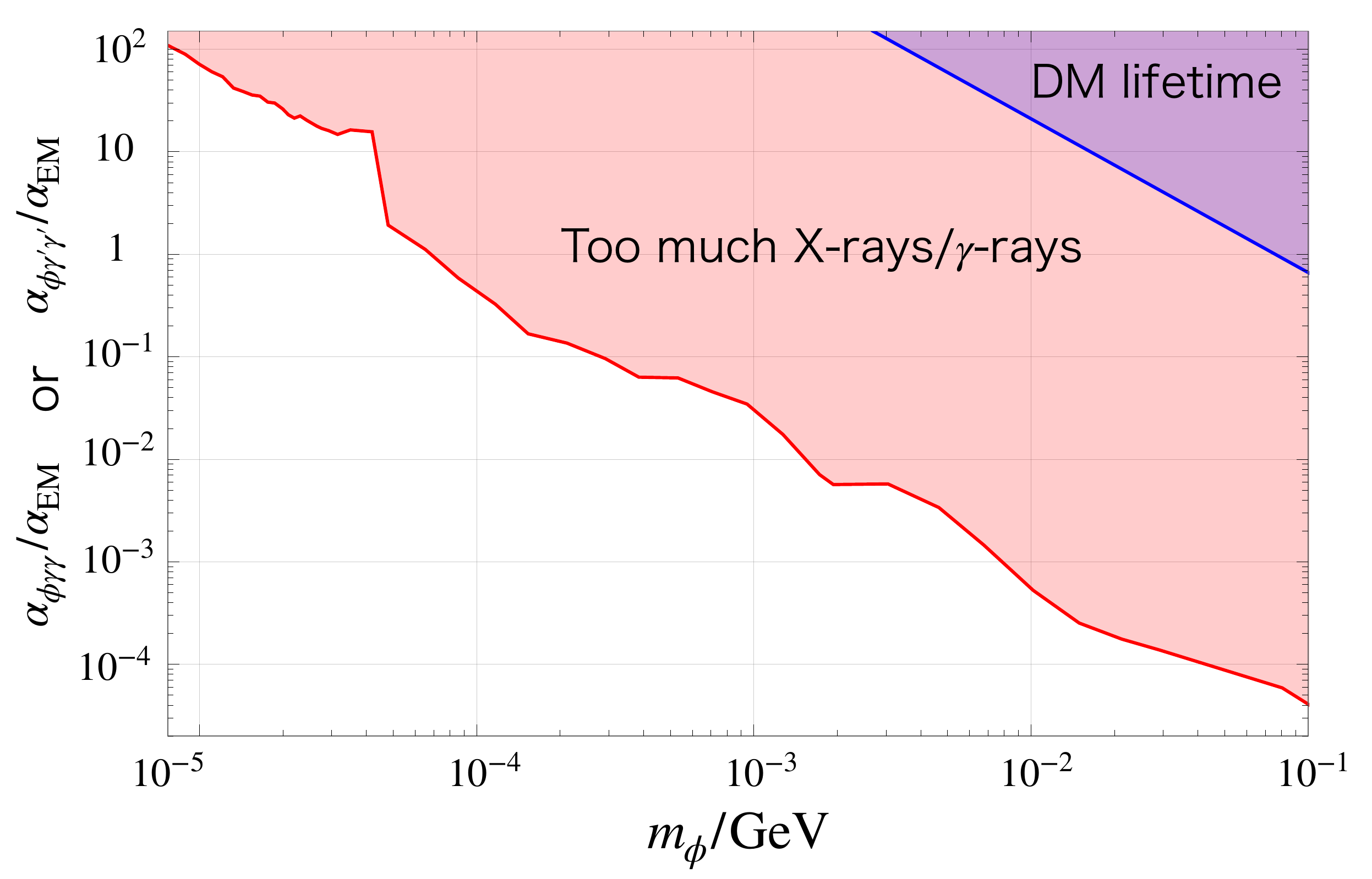}
\centering
\caption{The observational constraints on the axion coupling to photons and hidden photons.
The shaded regions are excluded, and we set $f_{\phi}=10^{16}\GeV$.
The red shaded region denotes the constraint (\ref{condition}) from the axion decay into photons,
and the blue region is the constraint (\ref{lifetime}) from the invisible decay of axion dark matter.
}
\label{diffuse}
\end{figure}

The constraint is placed on the parameters by the fact that the two contributions (\ref{MW}) and (\ref{EG}) should be smaller than the observational flux $F_{\rm{obs}}$. In this work, we compare the values integrated in the range of each energy bin width $\Delta E_\gamma$ with the observed flux,
\begin{eqnarray}
\int_{\Delta E_\gamma}dE_\gamma\left[\frac{1}{\Delta\Omega}E_\gamma^2\frac{d\Phi_{\rm{MW}}}{dE_{\gamma}}+E_\gamma^2\frac{d^2\Phi_{\rm{EG}}}{d\Omega dE_\gamma}\right]\lesssim F_{\rm{obs}}\Delta E_\gamma,
\label{condition}
\end{eqnarray}
where the unit $[\rm{MeVcm^{-2}s^{-1}sr^{-1}}]$ of the photon flux is used in our analysis.  We used the observational results from HEAO-1, INTEGRAL, COMPTEL, and EGRET summarized in~\cite{Essig:2013goa}. The results are shown in Fig.~\ref{diffuse} where we set  $f_{\phi}=10^{16}\GeV$. 
The red region represents the constraint (\ref{condition}) from the diffuse photon spectrum from the axion decay. The blue region is the constraint (\ref{lifetime}) from the dark matter lifetime. For $\alpha_{\phi\gamma\gamma}=\alpha_{\rm{EM}}$, the axion mass has an upper bound $m_{\phi}\lesssim70\KeV$.

%%%%%%%%%%%%%%%%%%%%%%%%%%%%%%%%%
\subsection{Primordial tensor mode
\label{sec:tensor}}
%%%%%%%%%%%%%%%%%%%%%%%%%%%%%%%%%
During inflation the primordial tensor mode, i.e. primordial gravitational waves, are
 generated, which contribute to the CMB temperature and polarization anisotropies. 
  Its amplitude is proportional to the inflation scale $H_{\rm inf}$. The absence
  of the primordial tensor mode so far in
the CMB observations by Planck and BICEP2/Keck Array provides an 
 upper bound on  $H_{\rm{inf}}$~\cite{Akrami:2018odb}: 
\beq
H_{\rm{inf}}<6.5\times10^{13}\GeV\hspace{10mm}(95\%~\rm{C.L.}).
\label{tensor}
\eeq

%%%%%%%%%%%%%%%%%%%%%%%%%%%%%%%%%
\subsection{Results
\label{sec:results}}
\begin{figure}[t!]
\includegraphics[width=13cm]{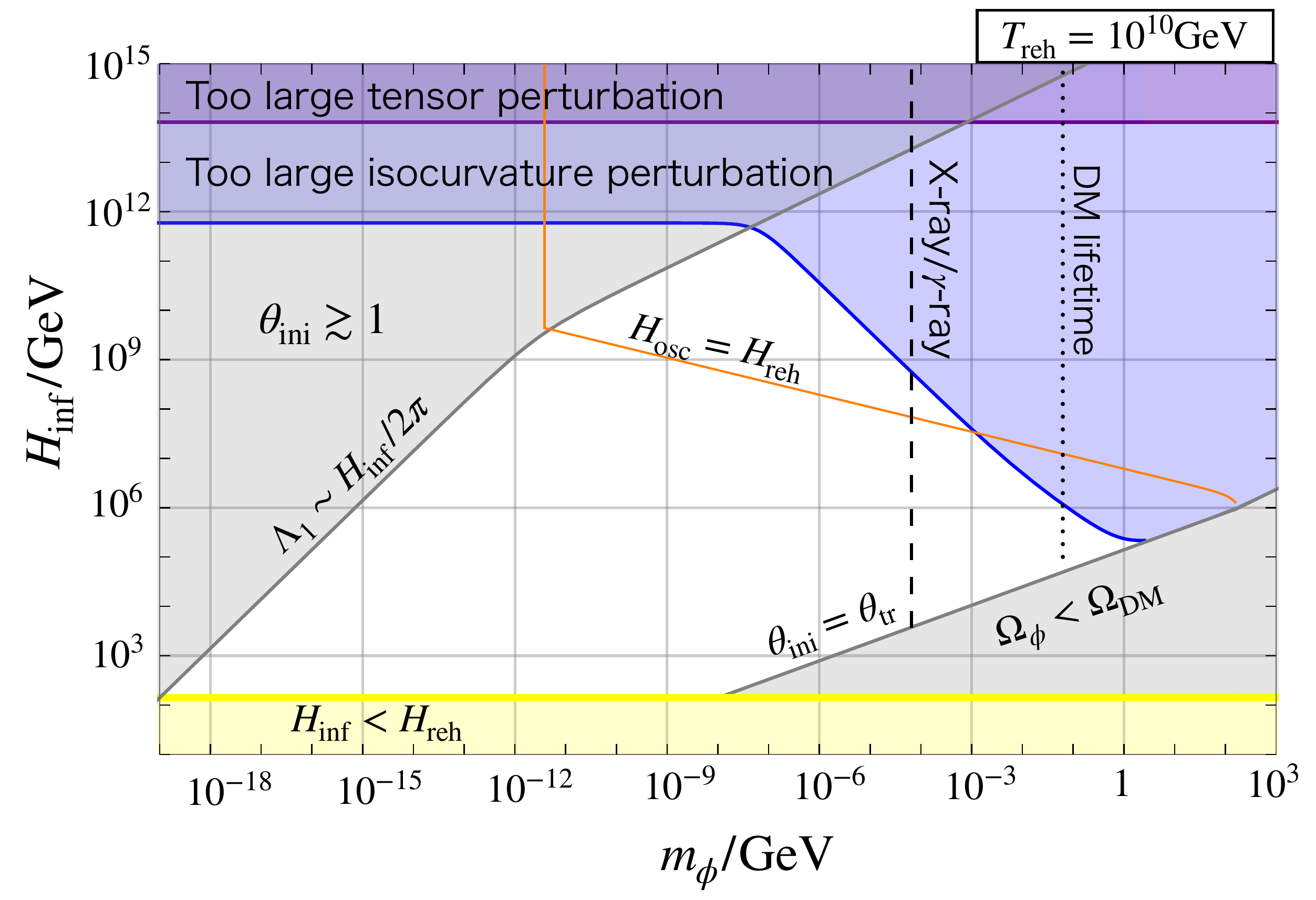}
\centering
\caption{The viable parameter space shown as a white region where $\Omega_{\phi}=\Omega_{\rm{DM}}$. We set $f_{\phi}=10^{16}\GeV$, and $T_{\rm{reh}}=10^{10}\GeV$. The lower-right gray line shows $\theta_{\rm{ini}}=\theta_{\rm{tr}}$ below which the axion abundance
falls short of the observed dark matter abundance. The purple and blue regions are the bounds from the absence of
 the primordial tensor mode (\ref{tensor}) and the isocurvature perturbation (\ref{isoconstrBD3}) and (\ref{isoconstrBD4}), respectively. The yellow region does not satisfy the condition that the reheating follows the inflation. The dotted vertical line is the constraint from the axion decay into hidden photons for $\alpha_{\phi\gamma'\gamma'}=\alpha_{\rm{EM}}$ and the dashed line is the constraint from the X-ray/$\gamma$-ray diffuse flux for $\alpha_{\phi\gamma\gamma}=\alpha_{\rm{EM}}$. The upper left gray region represents  $\theta_{\rm{ini}}\gtrsim1$. The orange solid line is the boundary $H_{\rm{reh}}=H_{\rm{osc}}$.  Note that the quadratic case corresponds to the lower right gray region.}
\label{para}
\end{figure}

\begin{figure}[t!]
\includegraphics[width=13cm]{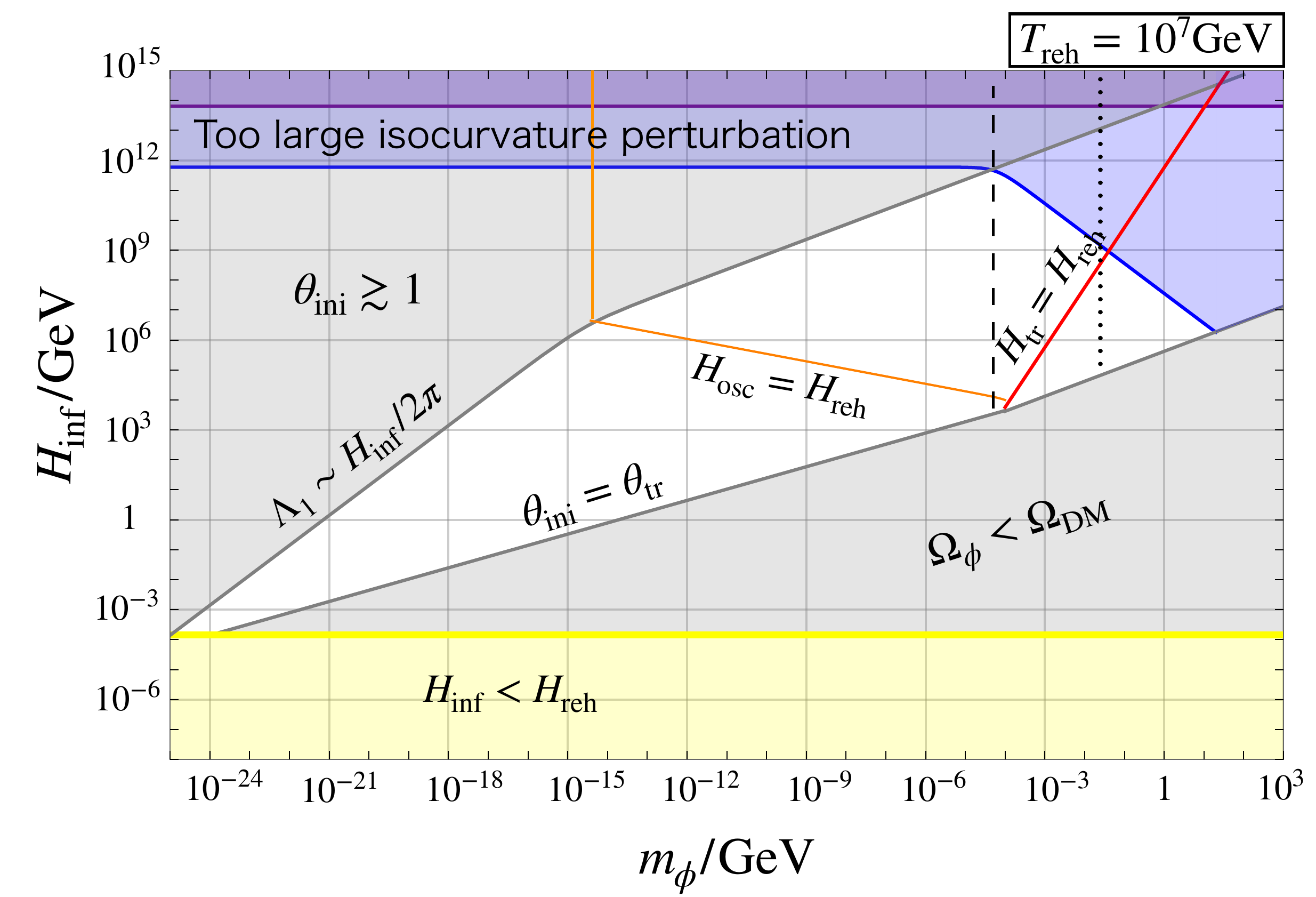}\\
\centering
\includegraphics[width=13cm]{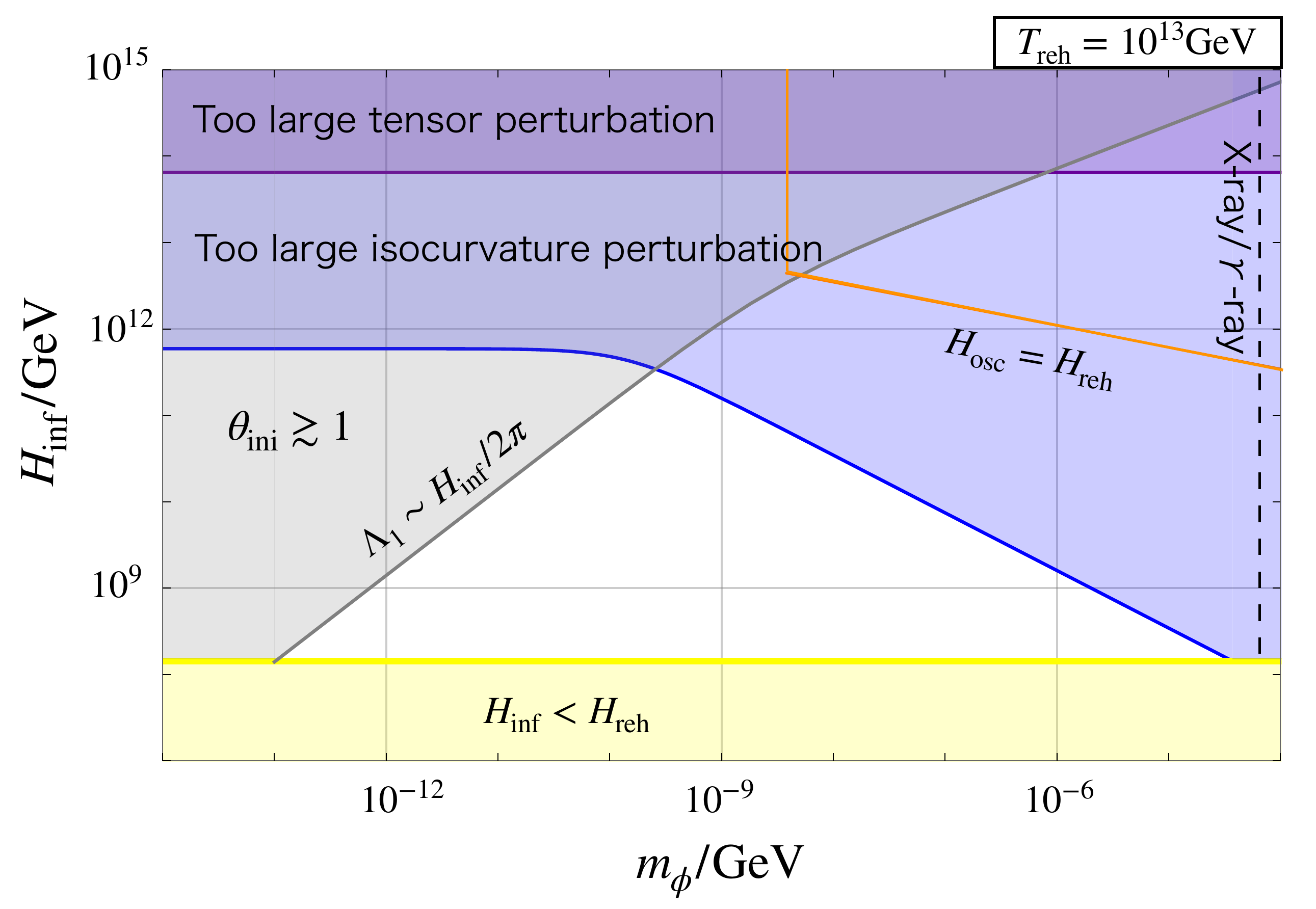}
\caption{
Same as Fig.~\ref{para} but for $T_{\rm{reh}}=10^{7}\GeV$ (upper) and $T_{\rm{reh}}=10^{13}\GeV$ (lower).
In the upper panel, the red line denotes $H_{\rm{reh}}=H_{\rm{tr}}$.
}
\label{result}
\end{figure}

%%%%%%%%%%%%%%%%%%%%%%%%%%%%%%%%%
Combining the above cosmological bounds, we show in Fig.~\ref{para} the viable parameter space  on the ($m_\phi$, $H_{\rm{inf}}$) plane
as a white region where the axion explains all dark matter. Here we fix $f_{\phi}=10^{16}\GeV$ and $T_{\rm{reh}}=10^{10}\GeV$,
and choose an appropriate value of $\Lambda_1$ for each  $m_\phi$ and $H_{\rm{inf}}$ to  realize 
$\Omega_\phi=\Omega_{\rm{DM}}$.
In  the lower right gray region,  $\theta_{\rm{ini}}$ is smaller than $\theta_{\rm{tr}}$, and the axion dynamics is same as for the 
quadratic  potential, and the axion abundance  is independent of $\Lambda_1$.
On the boundary line, the axion explains all dark matter,
while the axion abundance is smaller than the observed dark matter abundance below the gray line.
The purple and blue shaded regions are excluded by the absence of the tensor and isocurvature perturbations (\ref{tensor}),
(\ref{isoconstrBD3}) and (\ref{isoconstrBD4}), respectively.
The yellow shaded region at the bottom is excluded because of  $H_{\rm reh}>H_{\rm inf}.$ 
The dotted vertical line represents a constraint $m_{\phi}\lesssim80\MeV$, which comes from the invisible axion decay into 
hidden photons for $\alpha_{\phi\gamma'\gamma'}=\alpha_{\rm{EM}}$, while the dashed line  represents the constraint $m_{\phi}\lesssim70\KeV$ from the X-ray/$\gamma$-ray diffuse flux for $\alpha_{\phi\gamma\gamma}=\alpha_{\rm{EM}}$. 
In the upper left gray region,  $\theta_{\rm{ini}}$ given in Eq.~(\ref{initial2}) would become larger than unity, and 
our approximated potential is no longer justified. The isocurvature bound in this region is calculated assuming $\theta_{\rm{ini}}=1$.
Above the orange solid line, the axion starts to oscillate before the reheating, i.e., $H_{\rm{osc}} > H_{\rm{reh}}$.
The cases for $T_{\rm{reh}} = 10^7$ and $10^{13}$\,GeV are also shown in Fig.~\ref{result}. In the upper panel with $T_{\rm{reh}}=10^7\GeV$,
the red line denotes $H_{\rm{reh}}=H_{\rm{tr}}$. In the right to the red line, the reheating takes  place after the oscillation amplitude becomes
smaller than $\phi_{\rm tr}$,  $H_{\rm{reh}}\lesssim H_{\rm{tr}}\lesssim H_{\rm{osc}}$.

As mentioned above, in the quadratic case, a right amount of dark matter can be explained 
on the boundary line of the lower right gray region. The allowed region shown as a becomes much larger
in the case of the quartic potential.
This is because, for a given inflation scale,  the initial angle $\theta_{\rm{ini}}$ becomes smaller and the axion abundance behaves like
radiation and therefore decreases faster in the quartic case than in the quadratic case. 
In particular, the reheating temperature can be as high as $T_{\rm{reh}}\lesssim10^{14}\GeV$.  
 Thermal leptogenesis works for such a high reheating temperature ~\cite{Fukugita:1986hr}. 
For some range of the axion mass, the upper bound on the inflation scale is given by the isocurvature bound,
and so,  future CMB observations will be able to probe the corresponding parameter space. In addition, the relatively heavy axion dark matter
with mass  $m_\phi\sim10^{-6} - 1\GeV$ can be realized by virtue of the BD distribution. Therefore, if the axion couples to two photons,
next-generation gamma-ray satellites such as AMEGO~\cite{Moiseev:2017mxg, McEnery:2019tcm} and eASTROGAM~\cite{DeAngelis:2017gra}
will improve the bound significantly.
If the axion mainly couples to the hidden photons (or other hidden light particles), on the other hand, the present scenario may be probed
by further observations of the present and late-time universe (cf. Refs.~\cite{Enqvist:2015ara, Enqvist:2019tsa}).
In particular, the decaying dark matter may ameliorate the $H_0$ tension~\cite{Riess:2018byc}.

%%%%%%%%%%%%%%%%%%%%%%%%%%%%%%%%%%%%%%%%%%%
\section{Delayed onset of oscillations and $\pi$\hspace{-0.2mm}nflation
\label{sec:5}
}
%%%%%%%%%%%%%%%%%%%%%%%%%%%%%%%%%%%%%%%%%%%
So far we have focused on a case in which the axion mass at the minimum is
suppressed compared to the typical curvature scale of the potential. In this section we consider
another possibility that the axion potential has a plateau region away from the minimum which
delays the onset of oscillation. Such a possibility was  extensively studied in the
literature, because coherent oscillations in a potential flatter than the quadratic one leads 
to spatial instabilities, formation of non-topological solitons such as 
oscillons/I-balls~\cite{Gleiser:1993pt,Copeland:1995fq,Kasuya:2002zs} (see also Refs.~\cite{
McDonald:2001iv,Amin:2010jq,Amin:2011hj,Amin:2013ika,Takeda:2014qma,Kawasaki:2015vga,Lozanov:2016hid,
Hasegawa:2017iay,Antusch:2017flz,Hong:2017ooe,Ibe:2019vyo,Arvanitaki:2019rax} for more details about the
 formation and decay processes  in various contexts) 
 and the production of the gravitational waves~\cite{Zhou:2013tsa,Antusch:2016con,Kitajima:2018zco}. 
Also, the delayed onset of oscillations enhances the final axion abundance as well as
its isocurvature perturbation (see the footnote~\ref{ftnt}.)
The enhancement of  the axion (or ALP) abundance can also be obtained by e.g. the adiabatic conversion between the axion and the QCD axion~\cite{Kitajima:2014xla,Daido:2015bva,Daido:2015cba,Ho:2018qur}.

Here we present a model in which the initial value of the axion is set on the plateau of
the potential in the stochastic axion scenario. If we consider only the dynamics of the light axion, 
the inflation scale $H_{\rm inf}$ must be
higher than the height of the plateau, since otherwise the probability distribution of the axion would
be peaked at the potential minimum and it is unlikely to find the axion on the plateau of the potential. 
However, if there is another heavy axion field that mixes with the light axion, the probability distribution
of the lighter one can be shifted due to the heavy axion dynamics. Such a phase shift was recently
used to realize a hilltop initial condition for the QCD axion~\cite{Daido:2017wwb,Takahashi:2019pqf}.
(see also Ref.~\cite{Co:2018mho} which uses the stronger QCD in the early Universe~\cite{Dvali:1995ce,Banks:1996ea,Choi:1996fs,Jeong:2013xta}).

Let us introduce another heavy axion field, $\phi_H$, that mixes with the light axion. This is naturally
realized  in the axion lanscape. We also assume that the location of the local maxima and minima of
the heavy axion potential is the same as (or very close to) that of a single cosine term. 
For instance, this is the case if the heavy axion potential is dominated by a single cosine term, and another 
relatively small shift-symmetry breaking term(s) lifts the degeneracy of the vacua. Then, the universe may be 
first trapped in a false vacuum, and then tunnels to the adjacent lower vacuum, resulting in a field change by 
approximately $2\pi$~\cite{Daido:2017wwb}. 
Alternatively, if the heavy axion plays a role of the inflaton, we need at least two cosine terms as long as we consider
decay constants smaller than the Planck mass. Their relative height and phase are chosen so that the second
cosine term makes the potential maximum of the first one sufficiently flat for successful slow-roll inflation to occur.
Then, the field distance between the potential maximum and minimum is given by (a fraction of) $\pi$. 
This is due to the requirement of successful hilltop inflation. In either case, by introducing a certain mixing with the
light axion, one can realize a phase shift of $\pi$. In particular, such inflation model that causes the phase shift
of $\pi$ was named $\pi$\hspace{-0.2mm}nflation~\cite{Takahashi:2019pqf}.

To be concrete, let us consider the following potential for the light and
heavy axions,
\begin{align}
V_L(\phi,\theta_H)=\Lambda_1^4\left[1+\cos\left(\frac{\phi}{f_\phi}+ \theta_H \right)\right]+
\Lambda_2^4\left[1-\cos\left(2\left(\frac{\phi}{f_\phi}+\theta_H\right)\right)\right],
\end{align}
where we assume $\Lambda_1^4\simeq 4\Lambda_2^4$ as before, and $\theta_H \equiv \phi_H/f_H$ with $f_H$ being the decay constant  is the dimensionless 
angle corresponding to the heavy axion. We assume that the heavy axion has its own potential satisfying the
above mentioned property so that $\theta_H$ changes from $0$ during inflation to $\pi$ after inflation. 
We also assume that back reaction of $V_L(\phi, \theta_{\rm inf})$ to the heavy axion dynamics is negligibly
small.

During inflation the potential for $\phi$ is well approximated by the quadratic term
around one of the minima, $\phi = \pi f_\phi$,
\begin{align}
V_L(\phi,0)\simeq \frac{1}{2}M_{\phi}^2(\phi - \pi f_\phi)^2~~~~(\phi \sim \pi f_\phi).
\end{align}
where $M_\phi^2\simeq (\Lambda_1^4+4\Lambda_2^4)/f^2_\phi$ is the effective mass during inflation.
We assume $M_\phi \ll H_{\rm inf} \ll \Lambda_1$ and the BD distribution of $\phi$ is reached after a sufficiently long inflation.
The BD distribution of $\phi$ implies that $|\phi - \pi f_\phi|\sim H_{\rm inf}^2/M_\phi$, and it is peaked at $\phi = \pi f_\phi$.
 
 After inflation only the first term of the axion potential $V_L$ receives the phase shift of $\pi$, since the second one remains
 the same after the phase shift of $2\pi$. Then, the potential has a plateau around $\phi = \pi f_\phi$, and the minimum
is shifted to the origin. If the typical time scale  of the heavy axion dynamics is much shorter than $M_\phi^{-1}$, the
distribution  of the light axion remains almost the intact, and it is still peaked at $\phi = \pi f_\phi$. The potential can  be
expanded again around $\phi = \pi f_\phi$ which is now the potential maximum,
\begin{align}
V_L(\phi,\pi) \simeq 2\Lambda_1^4-\frac{1}{2}m_{\phi}^2(\phi-\pi f_\phi)^2-\frac{\lambda}{4!}(\phi-\pi f_\phi)^4
+ \cdots~~~~(\phi \sim \pi f_\phi)
\end{align}
On the other hand, the potential at around the minimum, $\phi = 0$, is approximately given by
\begin{align}
V_L(\phi,\pi)\simeq \frac{1}{2}M_{\phi}^2\phi^2 ~~~~(\phi \sim 0).
\end{align} 

For simplicity let us assume that the axion is initially in the vicinity of $\phi =  \pi f_\phi$ where the potential
is dominated by the (negative) quadratic term,  i.e., $|\theta_{\rm ini}-\pi|\ll \theta_{\rm tr}$ (see Eq.~(\ref{transition})
for the definition of $\theta_{\rm tr}$), and that the reheating occurs instantaneously just after inflation, 
i.e., $H_{\rm inf} \simeq H_{\rm reh}>H_{\rm osc}$. 
Then, one can determine the onset of oscillations from Eq.\,(\refeq{int}),
 \begin{equation}
H_{\rm osc}\sim  {\sqrt{\frac{\lambda}{60}}\theta_{\rm tr}}f_\phi,
 \end{equation}
 where we have used $c_{\rm osc} = 5$, and 
 we have neglected the logarithmic contributions from the quadratic terms for simplicity. 
 One finds that $H_{\rm osc}$ is not sensitive to $\theta_{\rm ini}$, and it is of order $m_\phi$.
After the axion starts to oscillate, the oscillation amplitude becomes smaller due to the cosmic expansion, and
its oscillation frequency is determined  by  $M_\phi.$ One gets the abundance of the axion by taking $\Lambda_1^4\simeq 4\Lambda_2^4$,
 \begin{equation}
\Omega_\phi h^2\sim  
0.5 \lrfp{M_\phi}{0.1 {\rm\,eV}}{\frac{1}{4}} \lrfp{f_\phi}{10^6{\rm\,GeV}}{\frac{7}{4}}
\left(\frac{\theta_{\rm tr}}{10^{-7}}\right)^{-3/2}.  
\end{equation} 
Consequently we get the light axion dark matter with a very small decay constant. 
The onset of the oscillation happens at $T\sim 10\,$GeV and the potential height is $\Lambda_1 \simeq 80$\,MeV.
Here the small $\theta_{\rm tr}$ controls the flatness of the plateau near $\phi = \pi f_\phi$, and therefore,
the enhancement of the axion abundance.
To have $\theta_{\rm ini}< \theta_{\rm tr}\sim 10^{-7}$, the Hubble parameter should be smaller than ${\cal O}(1)$MeV.\footnote{Such low-scale inflation and successful reheating can be 
realized in the ALP inflation~\cite{Daido:2017wwb, Daido:2017tbr, Takahashi:2019qmh}.  }
The isocurvature bound as well as the other constraints  can be satisfied. 
Interestingly, such a light axion dark matter with $M_\phi \sim 0.01-1\,$eV and $f_\phi\sim 10^6-10^7$\,GeV can be searched 
 in the IAXO experiment~\cite{Irastorza:2011gs, Armengaud:2014gea, Armengaud:2019uso} or the PTOLEMY experiment~\cite{McKeen:2018xyz, Chacko:2018uke, Betti:2019ouf} if the the axion couples to photons or neutrinos.

%%%%%%%%%%%%%%%%%%%%%%%%%%%%%%%%%%%%%%%%%%%%%%%%%%%%%%%%%%%%%%%%

%%%%%%%%%%%%%%%%%%%%%%%%%%%%%%%%%%%%%%%%%%%%%%%%%%%%%%%%%%%%%%%%
\section{Conclusions}
%%%%%%%%%%%%%%%%%%%%%%%%%%%%%%%%%%%%%%%%%%%%%%%%%%%%%%%%%%%%%
We have studied  the stochastic axion dark matter in the axion landscape that consists of many axions which generically have  mass and kinetic mixings. Because of the  various shift symmetry breaking  terms as well as mass mixings, 
we expect that some of the axion potential and its dynamics may not be captured by the usual analysis based on a single
cosine or a simple quadratic potential. In this paper, therefore, we have considered a possibility that the axion dynamics as well as
its abundance are significantly affected by the deviation from the simple quadratic potential. 

First, we have focused on a possibility that one of the light axions is responsible for the observed dark matter. In order to be sufficiently long-lived, the axion mass must be much lighter than the fundamental scale. 
This is realized either if there is a flat direction(s) in the landscape along which the typical curvature of the potential is very small,
or if the axion mass happens to be suppressed around the potential minimum. In the latter case, the potential is well approximated by the suppressed quadratic term plus a quartic term. We have carefully studied the axion initial condition determined by the stochastic dynamics and  its subsequent evolution after inflation. By taking account of various cosmological bounds, we have delineated  the viable parameter region in Figs.~\ref{para} and \ref{result}. We have found that the axion dark matter can be realized for a broad region of the inflation scale and the axion mass; e.g. $H_{\rm inf} = {\cal O}(10^{2-12})$\,GeV
and $m_\phi = 10^{-19} - 1$ GeV for $f_\phi = 10^{16}$\,GeV and $T_{\rm reh} = 10^{10}$\,GeV. 

We also discussed a case where the axion potential has a plateau away from the minimum, which
delays the onset of oscillations. In particular we have shown that one can realize the axion probability
distribution peaked at the plateau region by using the phase shift of $\pi$ induced by the heavy axion
dynamics. This is the so-called $\pi$\hspace{-0.2mm}nflation mechanism. Since the axion abundance can be
significantly enhanced in this case, one can explain the observed dark matter density for a very small $f_\phi$,
e.g. $f_\phi = 10^6$\,GeV and the axion mass of ${\cal O}(0.1-1)$\,eV. Such axion dark matter  with a small
$f_\phi$ may be probed by the IAXO or PTOLEMY experiments  if it has a coupling  to photons or neutrinos.

%---------------SECTION------------------%
%
\section*{Acknowledgments}
This work is supported by JSPS KAKENHI Grant Numbers
JP15H05889 (F.T.), JP15K21733 (F.T.),  JP17H02875 (F.T.), 
JP17H02878 (F.T.),  by World Premier International Research Center Initiative (WPI Initiative), MEXT, Japan, and by NRF Strategic Research Program NRF-2017R1E1A1A01072736 (W.Y.). One of us (S.N.) acknowledges support from the Graduate Program on Physics for the Universe (GP-PU) at Tohoku University.
%
%---------------SECTION------------------%

\vspace{1cm}

\bibliography{reference1}

\providecommand{\href}[2]{#2}\begingroup\raggedright\begin{thebibliography}{10}

\bibitem{Aghanim:2018eyx}
{\scshape Planck} collaboration, N.~Aghanim et~al., \emph{{Planck 2018 results.
  VI. Cosmological parameters}},
  \href{https://arxiv.org/abs/1807.06209}{{\ttfamily 1807.06209}}.

\bibitem{Preskill:1982cy}
J.~Preskill, M.~B. Wise and F.~Wilczek, \emph{{Cosmology of the Invisible
  Axion}}, \href{https://doi.org/10.1016/0370-2693(83)90637-8}{\emph{Phys.
  Lett.} {\bfseries 120B} (1983) 127}.

\bibitem{Abbott:1982af}
L.~F. Abbott and P.~Sikivie, \emph{{A Cosmological Bound on the Invisible
  Axion}}, \href{https://doi.org/10.1016/0370-2693(83)90638-X}{\emph{Phys.
  Lett.} {\bfseries 120B} (1983) 133}.

\bibitem{Dine:1982ah}
M.~Dine and W.~Fischler, \emph{{The Not So Harmless Axion}},
  \href{https://doi.org/10.1016/0370-2693(83)90639-1}{\emph{Phys. Lett.}
  {\bfseries 120B} (1983) 137}.

\bibitem{Graham:2018jyp}
P.~W. Graham and A.~Scherlis, \emph{{Stochastic axion scenario}},
  \href{https://doi.org/10.1103/PhysRevD.98.035017}{\emph{Phys. Rev.}
  {\bfseries D98} (2018) 035017}
  [\href{https://arxiv.org/abs/1805.07362}{{\ttfamily 1805.07362}}].

\bibitem{Guth:2018hsa}
F.~Takahashi, W.~Yin and A.~H. Guth, \emph{{QCD axion window and low-scale
  inflation}}, \href{https://doi.org/10.1103/PhysRevD.98.015042}{\emph{Phys.
  Rev.} {\bfseries D98} (2018) 015042}
  [\href{https://arxiv.org/abs/1805.08763}{{\ttfamily 1805.08763}}].

\bibitem{Bunch:1978yq}
T.~S. Bunch and P.~C.~W. Davies, \emph{{Quantum Field Theory in de Sitter
  Space: Renormalization by Point Splitting}},
  \href{https://doi.org/10.1098/rspa.1978.0060}{\emph{Proc. Roy. Soc. Lond.}
  {\bfseries A360} (1978) 117}.

\bibitem{Ho:2019ayl}
S.-Y. Ho, F.~Takahashi and W.~Yin, \emph{{Relaxing the Cosmological Moduli
  Problem by Low-scale Inflation}},
  \href{https://doi.org/10.1007/JHEP04(2019)149}{\emph{JHEP} {\bfseries 04}
  (2019) 149} [\href{https://arxiv.org/abs/1901.01240}{{\ttfamily
  1901.01240}}].

\bibitem{Higaki:2014pja}
T.~Higaki and F.~Takahashi, \emph{{Natural and Multi-Natural Inflation in Axion
  Landscape}}, \href{https://doi.org/10.1007/JHEP07(2014)074}{\emph{JHEP}
  {\bfseries 07} (2014) 074} [\href{https://arxiv.org/abs/1404.6923}{{\ttfamily
  1404.6923}}].

\bibitem{Higaki:2014mwa}
T.~Higaki and F.~Takahashi, \emph{{Axion Landscape and Natural Inflation}},
  \href{https://doi.org/10.1016/j.physletb.2015.03.052}{\emph{Phys. Lett.}
  {\bfseries B744} (2015) 153}
  [\href{https://arxiv.org/abs/1409.8409}{{\ttfamily 1409.8409}}].

\bibitem{Wang:2015rel}
G.~Wang and T.~Battefeld, \emph{{Vacuum Selection on Axionic Landscapes}},
  \href{https://doi.org/10.1088/1475-7516/2016/04/025}{\emph{JCAP} {\bfseries
  1604} (2016) 025} [\href{https://arxiv.org/abs/1512.04224}{{\ttfamily
  1512.04224}}].

\bibitem{Masoumi:2016eqo}
A.~Masoumi and A.~Vilenkin, \emph{{Vacuum statistics and stability in axionic
  landscapes}},
  \href{https://doi.org/10.1088/1475-7516/2016/03/054}{\emph{JCAP} {\bfseries
  1603} (2016) 054} [\href{https://arxiv.org/abs/1601.01662}{{\ttfamily
  1601.01662}}].

\bibitem{Nath:2017ihp}
P.~Nath and M.~Piskunov, \emph{{Evidence for Inflation in an Axion Landscape}},
  \href{https://doi.org/10.1007/JHEP03(2018)121}{\emph{JHEP} {\bfseries 03}
  (2018) 121} [\href{https://arxiv.org/abs/1712.01357}{{\ttfamily
  1712.01357}}].

\bibitem{Yamada:2017uzq}
M.~Yamada and A.~Vilenkin, \emph{{Hessian eigenvalue distribution in a random
  Gaussian landscape}},
  \href{https://doi.org/10.1007/JHEP03(2018)029}{\emph{JHEP} {\bfseries 03}
  (2018) 029} [\href{https://arxiv.org/abs/1712.01282}{{\ttfamily
  1712.01282}}].

\bibitem{Daido:2016tsj}
R.~Daido, T.~Kobayashi and F.~Takahashi, \emph{{Dark Matter in Axion
  Landscape}},
  \href{https://doi.org/10.1016/j.physletb.2016.12.034}{\emph{Phys. Lett.}
  {\bfseries B765} (2017) 293}
  [\href{https://arxiv.org/abs/1608.04092}{{\ttfamily 1608.04092}}].

\bibitem{Kim:2004rp}
J.~E. Kim, H.~P. Nilles and M.~Peloso, \emph{{Completing natural inflation}},
  \href{https://doi.org/10.1088/1475-7516/2005/01/005}{\emph{JCAP} {\bfseries
  0501} (2005) 005} [\href{https://arxiv.org/abs/hep-ph/0409138}{{\ttfamily
  hep-ph/0409138}}].

\bibitem{Choi:2014rja}
K.~Choi, H.~Kim and S.~Yun, \emph{{Natural inflation with multiple
  sub-Planckian axions}},
  \href{https://doi.org/10.1103/PhysRevD.90.023545}{\emph{Phys. Rev.}
  {\bfseries D90} (2014) 023545}
  [\href{https://arxiv.org/abs/1404.6209}{{\ttfamily 1404.6209}}].

\bibitem{Linde:1982ur}
A.~D. Linde, \emph{{NONSINGULAR REGENERATING INFLATIONARY UNIVERSE}}, .

\bibitem{Steinhardt:1982kg}
P.~J. Steinhardt, \emph{{NATURAL INFLATION}},  in \emph{{Nuffield Workshop on
  the Very Early Universe Cambridge, England, June 21-July 9, 1982}},
  pp.~251--266, 1982.

\bibitem{Vilenkin:1983xq}
A.~Vilenkin, \emph{{The Birth of Inflationary Universes}},
  \href{https://doi.org/10.1103/PhysRevD.27.2848}{\emph{Phys. Rev.} {\bfseries
  D27} (1983) 2848}.

\bibitem{Linde:1986fc}
A.~D. Linde, \emph{{ETERNAL CHAOTIC INFLATION}},
  \href{https://doi.org/10.1142/S0217732386000129}{\emph{Mod. Phys. Lett.}
  {\bfseries A1} (1986) 81}.

\bibitem{Linde:1986fd}
A.~D. Linde, \emph{{Eternally Existing Selfreproducing Chaotic Inflationary
  Universe}}, \href{https://doi.org/10.1016/0370-2693(86)90611-8}{\emph{Phys.
  Lett.} {\bfseries B175} (1986) 395}.

\bibitem{Goncharov:1987ir}
A.~S. Goncharov, A.~D. Linde and V.~F. Mukhanov, \emph{{The Global Structure of
  the Inflationary Universe}},
  \href{https://doi.org/10.1142/S0217751X87000211}{\emph{Int. J. Mod. Phys.}
  {\bfseries A2} (1987) 561}.

\bibitem{Guth:2000ka}
A.~H. Guth, \emph{{Inflation and eternal inflation}},
  \href{https://doi.org/10.1016/S0370-1573(00)00037-5}{\emph{Phys. Rept.}
  {\bfseries 333} (2000) 555}
  [\href{https://arxiv.org/abs/astro-ph/0002156}{{\ttfamily
  astro-ph/0002156}}].

\bibitem{Guth:2007ng}
A.~H. Guth, \emph{{Eternal inflation and its implications}},
  \href{https://doi.org/10.1088/1751-8113/40/25/S25}{\emph{J. Phys.} {\bfseries
  A40} (2007) 6811} [\href{https://arxiv.org/abs/hep-th/0702178}{{\ttfamily
  hep-th/0702178}}].

\bibitem{Linde:2015edk}
A.~Linde, \emph{{A brief history of the multiverse}},
  \href{https://doi.org/10.1088/1361-6633/aa50e4}{\emph{Rept. Prog. Phys.}
  {\bfseries 80} (2017) 022001}
  [\href{https://arxiv.org/abs/1512.01203}{{\ttfamily 1512.01203}}].

\bibitem{Kitajima:2019ibn}
N.~Kitajima, Y.~Tada and F.~Takahashi, \emph{{Stochastic inflation with an
  extremely large number of $e$-folds}},
  \href{https://doi.org/10.1016/j.physletb.2019.135097}{\emph{Phys. Lett.}
  {\bfseries B800} (2020) 135097}
  [\href{https://arxiv.org/abs/1908.08694}{{\ttfamily 1908.08694}}].

\bibitem{Gibbons:1977mu}
G.~W. Gibbons and S.~W. Hawking, \emph{{Cosmological Event Horizons,
  Thermodynamics, and Particle Creation}},
  \href{https://doi.org/10.1103/PhysRevD.15.2738}{\emph{Phys. Rev.} {\bfseries
  D15} (1977) 2738}.

\bibitem{Kobayashi:2013nva}
T.~Kobayashi, R.~Kurematsu and F.~Takahashi, \emph{{Isocurvature Constraints
  and Anharmonic Effects on QCD Axion Dark Matter}},
  \href{https://doi.org/10.1088/1475-7516/2013/09/032}{\emph{JCAP} {\bfseries
  1309} (2013) 032} [\href{https://arxiv.org/abs/1304.0922}{{\ttfamily
  1304.0922}}].

\bibitem{Kawasaki:2011pd}
M.~Kawasaki, T.~Kobayashi and F.~Takahashi, \emph{{Non-Gaussianity from
  Curvatons Revisited}}, \href{https://doi.org/10.1103/PhysRevD.84.123506,
  10.1103/PhysRevD.85.029905}{\emph{Phys. Rev.} {\bfseries D84} (2011) 123506}
  [\href{https://arxiv.org/abs/1107.6011}{{\ttfamily 1107.6011}}].

\bibitem{Salvio:2013iaa}
A.~Salvio, A.~Strumia and W.~Xue, \emph{{Thermal axion production}},
  \href{https://doi.org/10.1088/1475-7516/2014/01/011}{\emph{JCAP} {\bfseries
  1401} (2014) 011} [\href{https://arxiv.org/abs/1310.6982}{{\ttfamily
  1310.6982}}].

\bibitem{Akrami:2018odb}
{\scshape Planck} collaboration, Y.~Akrami et~al., \emph{{Planck 2018 results.
  X. Constraints on inflation}},
  \href{https://arxiv.org/abs/1807.06211}{{\ttfamily 1807.06211}}.

\bibitem{Enqvist:2015ara}
K.~Enqvist, S.~Nadathur, T.~Sekiguchi and T.~Takahashi, \emph{{Decaying dark
  matter and the tension in $\sigma_8$}},
  \href{https://doi.org/10.1088/1475-7516/2015/09/067}{\emph{JCAP} {\bfseries
  1509} (2015) 067} [\href{https://arxiv.org/abs/1505.05511}{{\ttfamily
  1505.05511}}].

\bibitem{Enqvist:2019tsa}
K.~Enqvist, S.~Nadathur, T.~Sekiguchi and T.~Takahashi, \emph{{Constraints on
  decaying dark matter from weak lensing and cluster counts}},
  \href{https://arxiv.org/abs/1906.09112}{{\ttfamily 1906.09112}}.

\bibitem{Essig:2013goa}
R.~Essig, E.~Kuflik, S.~D. McDermott, T.~Volansky and K.~M. Zurek,
  \emph{{Constraining Light Dark Matter with Diffuse X-Ray and Gamma-Ray
  Observations}}, \href{https://doi.org/10.1007/JHEP11(2013)193}{\emph{JHEP}
  {\bfseries 11} (2013) 193} [\href{https://arxiv.org/abs/1309.4091}{{\ttfamily
  1309.4091}}].

\bibitem{Navarro:1995iw}
J.~F. Navarro, C.~S. Frenk and S.~D.~M. White, \emph{{The Structure of cold
  dark matter halos}}, \href{https://doi.org/10.1086/177173}{\emph{Astrophys.
  J.} {\bfseries 462} (1996) 563}
  [\href{https://arxiv.org/abs/astro-ph/9508025}{{\ttfamily
  astro-ph/9508025}}].

\bibitem{Navarro:1996gj}
J.~F. Navarro, C.~S. Frenk and S.~D.~M. White, \emph{{A Universal density
  profile from hierarchical clustering}},
  \href{https://doi.org/10.1086/304888}{\emph{Astrophys. J.} {\bfseries 490}
  (1997) 493} [\href{https://arxiv.org/abs/astro-ph/9611107}{{\ttfamily
  astro-ph/9611107}}].

\bibitem{Asaka:1997rv}
T.~Asaka, J.~Hashiba, M.~Kawasaki and T.~Yanagida, \emph{{Cosmological moduli
  problem in gauge mediated supersymmetry breaking theories}},
  \href{https://doi.org/10.1103/PhysRevD.58.083509}{\emph{Phys. Rev.}
  {\bfseries D58} (1998) 083509}
  [\href{https://arxiv.org/abs/hep-ph/9711501}{{\ttfamily hep-ph/9711501}}].

\bibitem{Fukugita:1986hr}
M.~Fukugita and T.~Yanagida, \emph{{Baryogenesis Without Grand Unification}},
  \href{https://doi.org/10.1016/0370-2693(86)91126-3}{\emph{Phys. Lett.}
  {\bfseries B174} (1986) 45}.

\bibitem{Moiseev:2017mxg}
A.~Moiseev and O.~B. O. T.~A. Team, \emph{{All-Sky Medium Energy Gamma-ray
  Observatory (AMEGO)}}, \href{https://doi.org/10.22323/1.301.0798}{\emph{PoS}
  {\bfseries ICRC2017} (2018) 798}.

\bibitem{McEnery:2019tcm}
{\scshape AMEGO} collaboration, R.~Caputo et~al., \emph{{All-sky Medium Energy
  Gamma-ray Observatory: Exploring the Extreme Multimessenger Universe}},
  \href{https://arxiv.org/abs/1907.07558}{{\ttfamily 1907.07558}}.

\bibitem{DeAngelis:2017gra}
{\scshape e-ASTROGAM} collaboration, M.~Tavani et~al., \emph{{Science with
  e-ASTROGAM: A space mission for MeV-GeV gamma-ray astrophysics}},
  \href{https://doi.org/10.1016/j.jheap.2018.07.001}{\emph{JHEAp} {\bfseries
  19} (2018) 1} [\href{https://arxiv.org/abs/1711.01265}{{\ttfamily
  1711.01265}}].

\bibitem{Riess:2018byc}
A.~G. Riess et~al., \emph{{Milky Way Cepheid Standards for Measuring Cosmic
  Distances and Application to Gaia DR2: Implications for the Hubble
  Constant}}, \href{https://doi.org/10.3847/1538-4357/aac82e}{\emph{Astrophys.
  J.} {\bfseries 861} (2018) 126}
  [\href{https://arxiv.org/abs/1804.10655}{{\ttfamily 1804.10655}}].

\bibitem{Gleiser:1993pt}
M.~Gleiser, \emph{{Pseudostable bubbles}},
  \href{https://doi.org/10.1103/PhysRevD.49.2978}{\emph{Phys. Rev.} {\bfseries
  D49} (1994) 2978} [\href{https://arxiv.org/abs/hep-ph/9308279}{{\ttfamily
  hep-ph/9308279}}].

\bibitem{Copeland:1995fq}
E.~J. Copeland, M.~Gleiser and H.~R. Muller, \emph{{Oscillons: Resonant
  configurations during bubble collapse}},
  \href{https://doi.org/10.1103/PhysRevD.52.1920}{\emph{Phys. Rev.} {\bfseries
  D52} (1995) 1920} [\href{https://arxiv.org/abs/hep-ph/9503217}{{\ttfamily
  hep-ph/9503217}}].

\bibitem{Kasuya:2002zs}
S.~Kasuya, M.~Kawasaki and F.~Takahashi, \emph{{I-balls}},
  \href{https://doi.org/10.1016/S0370-2693(03)00344-7}{\emph{Phys. Lett.}
  {\bfseries B559} (2003) 99}
  [\href{https://arxiv.org/abs/hep-ph/0209358}{{\ttfamily hep-ph/0209358}}].

\bibitem{McDonald:2001iv}
J.~McDonald, \emph{{Inflaton condensate fragmentation in hybrid inflation
  models}}, \href{https://doi.org/10.1103/PhysRevD.66.043525}{\emph{Phys. Rev.}
  {\bfseries D66} (2002) 043525}
  [\href{https://arxiv.org/abs/hep-ph/0105235}{{\ttfamily hep-ph/0105235}}].

\bibitem{Amin:2010jq}
M.~A. Amin and D.~Shirokoff, \emph{{Flat-top oscillons in an expanding
  universe}}, \href{https://doi.org/10.1103/PhysRevD.81.085045}{\emph{Phys.
  Rev.} {\bfseries D81} (2010) 085045}
  [\href{https://arxiv.org/abs/1002.3380}{{\ttfamily 1002.3380}}].

\bibitem{Amin:2011hj}
M.~A. Amin, R.~Easther, H.~Finkel, R.~Flauger and M.~P. Hertzberg,
  \emph{{Oscillons After Inflation}},
  \href{https://doi.org/10.1103/PhysRevLett.108.241302}{\emph{Phys. Rev. Lett.}
  {\bfseries 108} (2012) 241302}
  [\href{https://arxiv.org/abs/1106.3335}{{\ttfamily 1106.3335}}].

\bibitem{Amin:2013ika}
M.~A. Amin, \emph{{K-oscillons: Oscillons with noncanonical kinetic terms}},
  \href{https://doi.org/10.1103/PhysRevD.87.123505}{\emph{Phys. Rev.}
  {\bfseries D87} (2013) 123505}
  [\href{https://arxiv.org/abs/1303.1102}{{\ttfamily 1303.1102}}].

\bibitem{Takeda:2014qma}
N.~Takeda and Y.~Watanabe, \emph{{No quasistable scalaron lump forms after
  $R^2$ inflation}},
  \href{https://doi.org/10.1103/PhysRevD.90.023519}{\emph{Phys. Rev.}
  {\bfseries D90} (2014) 023519}
  [\href{https://arxiv.org/abs/1405.3830}{{\ttfamily 1405.3830}}].

\bibitem{Kawasaki:2015vga}
M.~Kawasaki, F.~Takahashi and N.~Takeda, \emph{{Adiabatic Invariance of
  Oscillons/I-balls}},
  \href{https://doi.org/10.1103/PhysRevD.92.105024}{\emph{Phys. Rev.}
  {\bfseries D92} (2015) 105024}
  [\href{https://arxiv.org/abs/1508.01028}{{\ttfamily 1508.01028}}].

\bibitem{Lozanov:2016hid}
K.~D. Lozanov and M.~A. Amin, \emph{{Equation of State and Duration to
  Radiation Domination after Inflation}},
  \href{https://doi.org/10.1103/PhysRevLett.119.061301}{\emph{Phys. Rev. Lett.}
  {\bfseries 119} (2017) 061301}
  [\href{https://arxiv.org/abs/1608.01213}{{\ttfamily 1608.01213}}].

\bibitem{Hasegawa:2017iay}
F.~Hasegawa and J.-P. Hong, \emph{{Inflaton fragmentation in E-models of
  cosmological $\alpha$-attractors}},
  \href{https://doi.org/10.1103/PhysRevD.97.083514}{\emph{Phys. Rev.}
  {\bfseries D97} (2018) 083514}
  [\href{https://arxiv.org/abs/1710.07487}{{\ttfamily 1710.07487}}].

\bibitem{Antusch:2017flz}
S.~Antusch, F.~Cefala, S.~Krippendorf, F.~Muia, S.~Orani and F.~Quevedo,
  \emph{{Oscillons from String Moduli}},
  \href{https://doi.org/10.1007/JHEP01(2018)083}{\emph{JHEP} {\bfseries 01}
  (2018) 083} [\href{https://arxiv.org/abs/1708.08922}{{\ttfamily
  1708.08922}}].

\bibitem{Hong:2017ooe}
J.-P. Hong, M.~Kawasaki and M.~Yamazaki, \emph{{Oscillons from Pure Natural
  Inflation}}, \href{https://doi.org/10.1103/PhysRevD.98.043531}{\emph{Phys.
  Rev.} {\bfseries D98} (2018) 043531}
  [\href{https://arxiv.org/abs/1711.10496}{{\ttfamily 1711.10496}}].

\bibitem{Ibe:2019vyo}
M.~Ibe, M.~Kawasaki, W.~Nakano and E.~Sonomoto, \emph{{Decay of I-ball/Oscillon
  in Classical Field Theory}},
  \href{https://doi.org/10.1007/JHEP04(2019)030}{\emph{JHEP} {\bfseries 04}
  (2019) 030} [\href{https://arxiv.org/abs/1901.06130}{{\ttfamily
  1901.06130}}].

\bibitem{Arvanitaki:2019rax}
A.~Arvanitaki, S.~Dimopoulos, M.~Galanis, L.~Lehner, J.~O. Thompson and
  K.~Van~Tilburg, \emph{{The Large-Misalignment Mechanism for the Formation of
  Compact Axion Structures: Signatures from the QCD Axion to Fuzzy Dark
  Matter}},  \href{https://arxiv.org/abs/1909.11665}{{\ttfamily 1909.11665}}.

\bibitem{Zhou:2013tsa}
S.-Y. Zhou, E.~J. Copeland, R.~Easther, H.~Finkel, Z.-G. Mou and P.~M. Saffin,
  \emph{{Gravitational Waves from Oscillon Preheating}},
  \href{https://doi.org/10.1007/JHEP10(2013)026}{\emph{JHEP} {\bfseries 10}
  (2013) 026} [\href{https://arxiv.org/abs/1304.6094}{{\ttfamily 1304.6094}}].

\bibitem{Antusch:2016con}
S.~Antusch, F.~Cefala and S.~Orani, \emph{{Gravitational waves from oscillons
  after inflation}}, \href{https://doi.org/10.1103/PhysRevLett.120.219901,
  10.1103/PhysRevLett.118.011303}{\emph{Phys. Rev. Lett.} {\bfseries 118}
  (2017) 011303} [\href{https://arxiv.org/abs/1607.01314}{{\ttfamily
  1607.01314}}].

\bibitem{Kitajima:2018zco}
N.~Kitajima, J.~Soda and Y.~Urakawa, \emph{{Gravitational wave forest from
  string axiverse}},
  \href{https://doi.org/10.1088/1475-7516/2018/10/008}{\emph{JCAP} {\bfseries
  1810} (2018) 008} [\href{https://arxiv.org/abs/1807.07037}{{\ttfamily
  1807.07037}}].

\bibitem{Kitajima:2014xla}
N.~Kitajima and F.~Takahashi, \emph{{Resonant conversions of QCD axions into
  hidden axions and suppressed isocurvature perturbations}},
  \href{https://doi.org/10.1088/1475-7516/2015/01/032}{\emph{JCAP} {\bfseries
  1501} (2015) 032} [\href{https://arxiv.org/abs/1411.2011}{{\ttfamily
  1411.2011}}].

\bibitem{Daido:2015bva}
R.~Daido, N.~Kitajima and F.~Takahashi, \emph{{Domain Wall Formation from Level
  Crossing in the Axiverse}},
  \href{https://doi.org/10.1103/PhysRevD.92.063512}{\emph{Phys. Rev.}
  {\bfseries D92} (2015) 063512}
  [\href{https://arxiv.org/abs/1505.07670}{{\ttfamily 1505.07670}}].

\bibitem{Daido:2015cba}
R.~Daido, N.~Kitajima and F.~Takahashi, \emph{{Level crossing between the QCD
  axion and an axionlike particle}},
  \href{https://doi.org/10.1103/PhysRevD.93.075027}{\emph{Phys. Rev.}
  {\bfseries D93} (2016) 075027}
  [\href{https://arxiv.org/abs/1510.06675}{{\ttfamily 1510.06675}}].

\bibitem{Ho:2018qur}
S.-Y. Ho, K.~Saikawa and F.~Takahashi, \emph{{Enhanced photon coupling of ALP
  dark matter adiabatically converted from the QCD axion}},
  \href{https://doi.org/10.1088/1475-7516/2018/10/042}{\emph{JCAP} {\bfseries
  1810} (2018) 042} [\href{https://arxiv.org/abs/1806.09551}{{\ttfamily
  1806.09551}}].

\bibitem{Daido:2017wwb}
R.~Daido, F.~Takahashi and W.~Yin, \emph{{The ALP miracle: unified inflaton and
  dark matter}},
  \href{https://doi.org/10.1088/1475-7516/2017/05/044}{\emph{JCAP} {\bfseries
  1705} (2017) 044} [\href{https://arxiv.org/abs/1702.03284}{{\ttfamily
  1702.03284}}].

\bibitem{Takahashi:2019pqf}
F.~Takahashi and W.~Yin, \emph{{QCD axion on hilltop by a phase shift of
  $\pi$}}, \href{https://doi.org/10.1007/JHEP10(2019)120}{\emph{JHEP}
  {\bfseries 10} (2019) 120}
  [\href{https://arxiv.org/abs/1908.06071}{{\ttfamily 1908.06071}}].

\bibitem{Co:2018mho}
R.~T. Co, E.~Gonzalez and K.~Harigaya, \emph{{Axion Misalignment Driven to the
  Hilltop}}, \href{https://doi.org/10.1007/JHEP05(2019)163}{\emph{JHEP}
  {\bfseries 05} (2019) 163}
  [\href{https://arxiv.org/abs/1812.11192}{{\ttfamily 1812.11192}}].

\bibitem{Dvali:1995ce}
G.~R. Dvali, \emph{{Removing the cosmological bound on the axion scale}},
  \href{https://arxiv.org/abs/hep-ph/9505253}{{\ttfamily hep-ph/9505253}}.

\bibitem{Banks:1996ea}
T.~Banks and M.~Dine, \emph{{The Cosmology of string theoretic axions}},
  \href{https://doi.org/10.1016/S0550-3213(97)00413-6}{\emph{Nucl. Phys.}
  {\bfseries B505} (1997) 445}
  [\href{https://arxiv.org/abs/hep-th/9608197}{{\ttfamily hep-th/9608197}}].

\bibitem{Choi:1996fs}
K.~Choi, H.~B. Kim and J.~E. Kim, \emph{{Axion cosmology with a stronger QCD in
  the early universe}},
  \href{https://doi.org/10.1016/S0550-3213(97)00066-7}{\emph{Nucl. Phys.}
  {\bfseries B490} (1997) 349}
  [\href{https://arxiv.org/abs/hep-ph/9606372}{{\ttfamily hep-ph/9606372}}].

\bibitem{Jeong:2013xta}
K.~S. Jeong and F.~Takahashi, \emph{{Suppressing Isocurvature Perturbations of
  QCD Axion Dark Matter}},
  \href{https://doi.org/10.1016/j.physletb.2013.10.061}{\emph{Phys. Lett.}
  {\bfseries B727} (2013) 448}
  [\href{https://arxiv.org/abs/1304.8131}{{\ttfamily 1304.8131}}].

\bibitem{Daido:2017tbr}
R.~Daido, F.~Takahashi and W.~Yin, \emph{{The ALP miracle revisited}},
  \href{https://doi.org/10.1007/JHEP02(2018)104}{\emph{JHEP} {\bfseries 02}
  (2018) 104} [\href{https://arxiv.org/abs/1710.11107}{{\ttfamily
  1710.11107}}].

\bibitem{Takahashi:2019qmh}
F.~Takahashi and W.~Yin, \emph{{ALP inflation and Big Bang on Earth}},
  \href{https://doi.org/10.1007/JHEP07(2019)095}{\emph{JHEP} {\bfseries 07}
  (2019) 095} [\href{https://arxiv.org/abs/1903.00462}{{\ttfamily
  1903.00462}}].

\bibitem{Irastorza:2011gs}
I.~G. Irastorza et~al., \emph{{Towards a new generation axion helioscope}},
  \href{https://doi.org/10.1088/1475-7516/2011/06/013}{\emph{JCAP} {\bfseries
  1106} (2011) 013} [\href{https://arxiv.org/abs/1103.5334}{{\ttfamily
  1103.5334}}].

\bibitem{Armengaud:2014gea}
E.~Armengaud et~al., \emph{{Conceptual Design of the International Axion
  Observatory (IAXO)}},
  \href{https://doi.org/10.1088/1748-0221/9/05/T05002}{\emph{JINST} {\bfseries
  9} (2014) T05002} [\href{https://arxiv.org/abs/1401.3233}{{\ttfamily
  1401.3233}}].

\bibitem{Armengaud:2019uso}
{\scshape IAXO} collaboration, E.~Armengaud et~al., \emph{{Physics potential of
  the International Axion Observatory (IAXO)}},
  \href{https://doi.org/10.1088/1475-7516/2019/06/047}{\emph{JCAP} {\bfseries
  1906} (2019) 047} [\href{https://arxiv.org/abs/1904.09155}{{\ttfamily
  1904.09155}}].

\bibitem{McKeen:2018xyz}
D.~McKeen, \emph{{Cosmic neutrino background search experiments as decaying
  dark matter detectors}},
  \href{https://doi.org/10.1103/PhysRevD.100.015028}{\emph{Phys. Rev.}
  {\bfseries D100} (2019) 015028}
  [\href{https://arxiv.org/abs/1812.08178}{{\ttfamily 1812.08178}}].

\bibitem{Chacko:2018uke}
Z.~Chacko, P.~Du and M.~Geller, \emph{{Detecting a Secondary Cosmic Neutrino
  Background from Majoron Decays in Neutrino Capture Experiments}},
  \href{https://doi.org/10.1103/PhysRevD.100.015050}{\emph{Phys. Rev.}
  {\bfseries D100} (2019) 015050}
  [\href{https://arxiv.org/abs/1812.11154}{{\ttfamily 1812.11154}}].

\bibitem{Betti:2019ouf}
{\scshape PTOLEMY} collaboration, M.~G. Betti et~al., \emph{{Neutrino physics
  with the PTOLEMY project: active neutrino properties and the light sterile
  case}}, \href{https://doi.org/10.1088/1475-7516/2019/07/047}{\emph{JCAP}
  {\bfseries 1907} (2019) 047}
  [\href{https://arxiv.org/abs/1902.05508}{{\ttfamily 1902.05508}}].

\end{thebibliography}\endgroup

\end{document}